\documentclass[longtitle,preprint,5p,times,twocolumn]{elsarticle}
\pdfoutput=1
\usepackage{amssymb}

\usepackage{caption}
\usepackage{xspace}
\usepackage{longtable}
\usepackage{pdflscape}
\usepackage{multirow}
\usepackage{ragged2e}
\usepackage{booktabs}
\usepackage{hyperref}
\usepackage{amsmath}
\usepackage[switch]{lineno}

\journal{Nuclear Instruments and Methods A}
\DeclareUnicodeCharacter{2212}{-}
\begin{document}
\begin{frontmatter}
\date{}
\title{Exclusive J/$\psi$ Detection and Physics with ECCE}
\def\theaffn{\arabic{affn}} 


%
%
%
%
\author[USTC]{X.~Li}
\author[MoreheadState]{J.~K.~Adkins}
\author[RIKEN,RBRC]{Y.~Akiba}
\author[UKansas]{A.~Albataineh}
\author[ODU]{M.~Amaryan}
\author[Oslo]{I.~C.~Arsene}
\author[MSU]{C. Ayerbe Gayoso}
\author[Sungkyunkwan]{J.~Bae}
\author[UVA]{X.~Bai}
\author[BNL,JLab]{M.D.~Baker}
% ...

%
\affiliation[USTC]{organization={University of Science and Technology of China},
     city={Hefei},
     country={China}}
     
% ...
\affiliation[BNL]{organization={Brookhaven National Laboratory},
	 city={Upton},
	 state={NY},
	 country={USA}} 
 
% ...
\affiliation[JLab]{organization={Thomas Jefferson National Accelerator Facility},
	 city={Newport News},
	 state={VA},
	 country={USA}} 
 
% ...
\affiliation[MoreheadState]{organization={Morehead State University},
	 city={Morehead},
	 state={KY},
	 }
 
% ...
\affiliation[MSU]{organization={Mississippi State University},
	 city={Mississippi State},
	 state={MS},
	 country={USA}} 
 
% ...
\affiliation[ODU]{organization={Old Dominion University},
	 city={Norfolk},
	 state={VA},
	 country={USA}} 
 
% ...
\affiliation[RIKEN]{organization={RIKEN Nishina Center},
	 city={Wako},
	 state={Saitama},
	 country={Japan}} 
 
% ...
\affiliation[RBRC]{organization={RIKEN BNL Research Center},
	 city={Upton},
	 state={NY},
	 country={USA}} 
	 
% ...
\affiliation[Sungkyunkwan]{organization={Sungkyunkwan University},
	 city={Suwon},
	 country={Republic of Korea}} 
 
% ...
\affiliation[UKansas]{organization={Unviersity of Kansas},
	 city={Lawrence},
	 state={KS},
	 country={USA}} 
 
% ...
\affiliation[Oslo]{organization={University of Oslo},
	 city={Oslo},
	 country={Norway}} 
 
% ...
\affiliation[UVA]{organization={University of Virginia},
	 city={Charlottesville},
	 state={VA},
	 country={USA}} 
 
% ...
	 \cortext[cor1]{first@ustc.edu.cn}

%
\begin{abstract}
Exclusive heavy quarkonium photoproduction is one of the most popular processes in EIC, which has a large cross section and a simple final state. Due to the gluonic nature of the exchange Pomeron, this process can be related to the gluon distributions in the nucleus. The momentum transfer dependence of this process is sensitive to the interaction sites, which provides a powerful tool to probe the spatial distribution of gluons in the nucleus. Recently the problem of the origin of hadron mass has received lots of attention in determining the anomaly contribution $M_a$. The trace anomaly is sensitive to the gluon condensate, and exclusive production of quarkonia such as J/$\psi$ and $\Upsilon$ can serve as a sensitive probe to constrain it. In this paper, we present the performance of the ECCE detector for exclusive J/$\psi$ detection and the capability of this process to investigate the above physics opportunities with ECCE. 
\end{abstract}



\end{frontmatter}
\setcounter{tocdepth}{1}
\tableofcontents

\section {Introduction}
\label{Introduction}
\begin{figure*}[!h]
    \centering
        \includegraphics[width=0.37\linewidth]{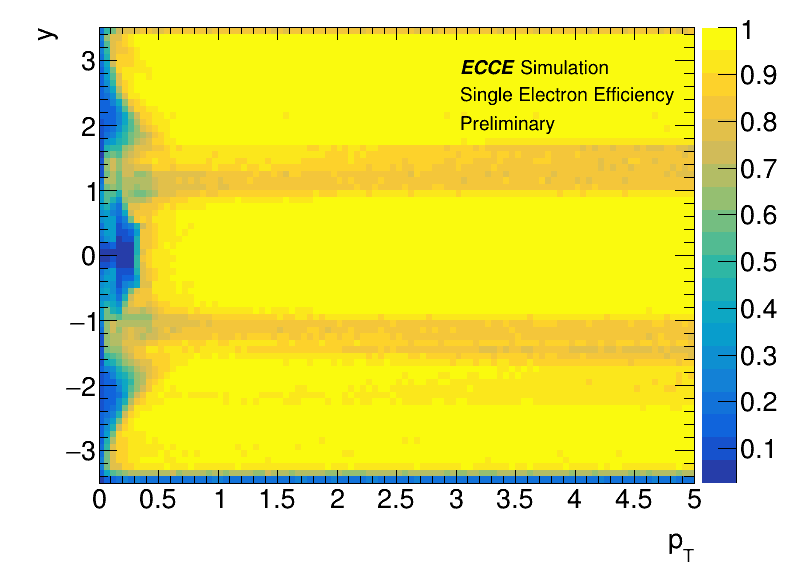}
        \includegraphics[width=0.37\linewidth]{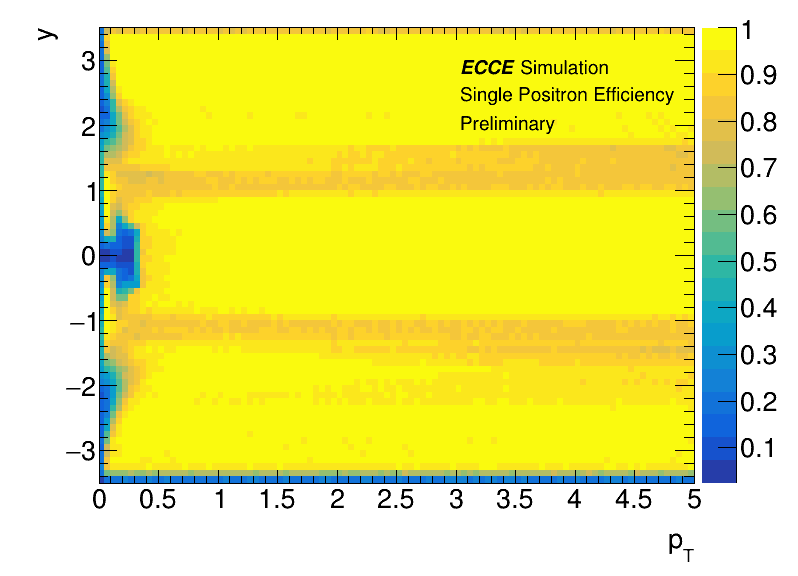}
    \caption{Single track efficiency. Left Panel: $e^-$ efficiency. Right Panel: $e^+$ efficiency}
    \label{fig:singletrack}
\end{figure*}
Nuclear parton distribution functions (nPDF) describe the behavior of bound partons in the nucleus. Most of the understanding of nPDF comes from fixed-target experiments. Determination of nPDF is through global fits to existing inclusive deep inelastic scattering (DIS) data. Constructing the ratio nPDF/PDF to quantified nuclear modifications is natural, which can cancel many of the theory uncertainties. A ratio below unity is called shadowing, while an enhancement is known as anti-shadowing. Recently a moderate gluon shadowing has been exhibited by J/$\psi$ photoproduction data from LHC~\cite{new1,new2,new3}. However, little is known about anti-shadowing at $x \sim 0.1$. The realization of the EIC with variable ion beam species will enable measurements of nPDF over a broad range of x and $Q^2$. Photoproduction of vector meson via photon-Pomeron fusion is able to cleanly and clearly determine nuclear gluon PDFs at the EIC. With broad x coverage, J/$\psi$ photoproduction can provide precise measurements to deepen our understanding of shadowing and anti-shadowing.\newline

Exclusive photoproduction, which has a large cross section and a simple final state, is projected to play a prominent role in the heavy quarkonia production processes at the EIC. In the reaction, a virtual incident photon fluctuates into a quark-antiquark pair, which scatters elastically off the target and emerges as a real quarkonium. The scattering process occurs via the exchange of a color neutral object, Pomeron, which can be viewed as two gluons with self interaction (gluon ladder) in the language of QCD. Due to the gluonic nature of Pomeron, the exclusive heavy quarkonia photoproduction at EIC can be related to the gluon distributions in the proton and nucleus using perturbative QCD. Furthermore, the distribution of momentum transfer from the target in the process is sensitive to the interaction sites, which provides a powerful tool to probe the spatial distribution of gluon in the nucleus. \newline

Nucleons constitute about 99$\%$ of the mass of the visible universe. In the standard model, Higgs mechanism describes gauge bosons' ``mass" generation. However it can only account for a small fraction of the nucleon mass. The major part comes from the strong interaction that binds quarks and gluons together. Understanding the hadron mass decomposition from strong interaction has become a topic of great interest in QCD. There are two key models~\cite{models1,models2,models3,models4,models5,models6} for the mass decomposition. One contains a trace anomaly contribution which is quantified by the energy-momentum tensor (EMT), and the other one agrees with an energy decomposition in the rest frame of the system. Recently, there has been sustained interest~\cite{Ji:1995sv,Ji:2021mtz,Guo:2021ibg} among the nucleon structure community in determining the anomaly contribution $M_a$ as a key to understanding the origin of the proton mass. Specifically, it has been proposed, based on some theorists' suggestions~\cite{oo2001,oo2016,oo2020}, that $M_a$ can be accessed through the forward (t=0) cross section via the exclusive production of heavy quarkonia states such as J/$\psi$ and $\Upsilon$. Heavy quarkonia are of particular interest here because they only couple to gluons, not to light quarks, and are thus sensitive to the gluonic structure of the proton. The trace anomaly is sensitive to the gluon condensate, with sensitivity greatest for production around the threshold.\\

In this paper, we simulate exclusive J/$\psi$ production using Fun4All framework with the designed ECCE detector system. In the simulation, we utilize eSTARLight model as the event generator for the exclusive photoproduction process. We make a projection of the exclusive J/$\psi$ measurement at ECCE under the designed integrated luminosity of one year running for EIC to give an insight into related fruitful physics opportunities, such as probing the nuclear gluon PDF, spatial distribution and proton mass decomposition. The major goal of this research is to present the detection capability and the physics opportunities which could be achieved with the ECCE detector setup for the exclusive process of J/$\psi$ photoproduction. 

%
%


\section{Simulation Framework of ECCE Detector Setup for J/$\psi$ Detection}
\label{setup}
The ECCE detector is a cylindrical detector covering $\left|\eta\right| \leq 3.5$ and the full azimuth. ECCE's tracking and vertexing systems use semiconductor and gaseous tracking detector technologies: Monolithic Active Pixel Sensor (MAPS) based silicon vertex/tracking detector and $\mu$Rwell based gas tracker derived from Gas Electron Multiplier (GEM) technology. 
 According to the simulation of the designed tracking system, the momentum resolution of the central region and beam e-going direction is closed to or better than the requirement of Yellow Report (YR)~\cite{khalek2021science}. 

For exclusive photoproduction of J/$\psi$, we adopt eSTARLight prediction of the cross section for $e p \rightarrow e J/\psi p$ process with two minor improvements, detailed in Sec.~\ref{theoretical:setup}. eSTARLight provides a photo-Pomeron interaction model parameterized by HERA data. In this study, two beam configurations, 5$\times$41 GeV and 10$\times$100 GeV, are used for e+p and e+Au collisions. 

The detector response simulation is done by a GEANT4 based package called Fun4All. In this work, the "Prop.7" detector concept is employed in J/$\psi$ reconstruction via dielectron channel. Single $e^+/e^-$ Tracking simulation results are shown as Fig.~\ref{fig:singletrack}. The difference in efficiency between $e^+$ and $e^-$ at very low $p_T$ is due to the initial assumption parameter in the Kalman filter. If the beginning parameter is set to ``positron," negative charge particles will have a low match quality and will likely be rejected. 


\begin{figure}
    \centering
        \includegraphics[width=0.8\linewidth]{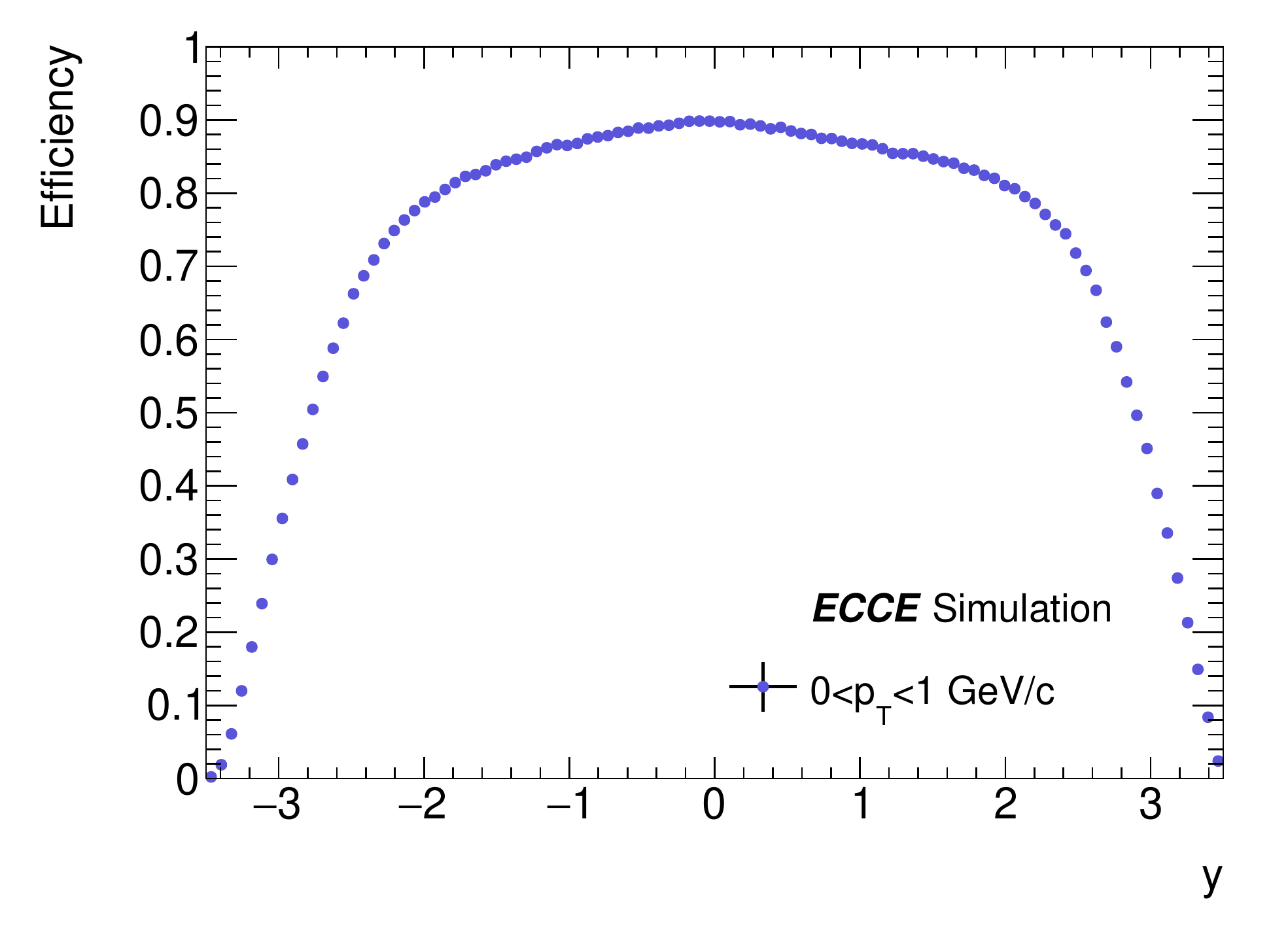}
    \caption{Tracking efficiency of J/$\psi$ from exclusive J/$\psi$ simulation.}
    \label{fig:singlejpsi}
\end{figure}

\begin{figure*}[!h]
    \centering
        \includegraphics[width=0.37\linewidth]{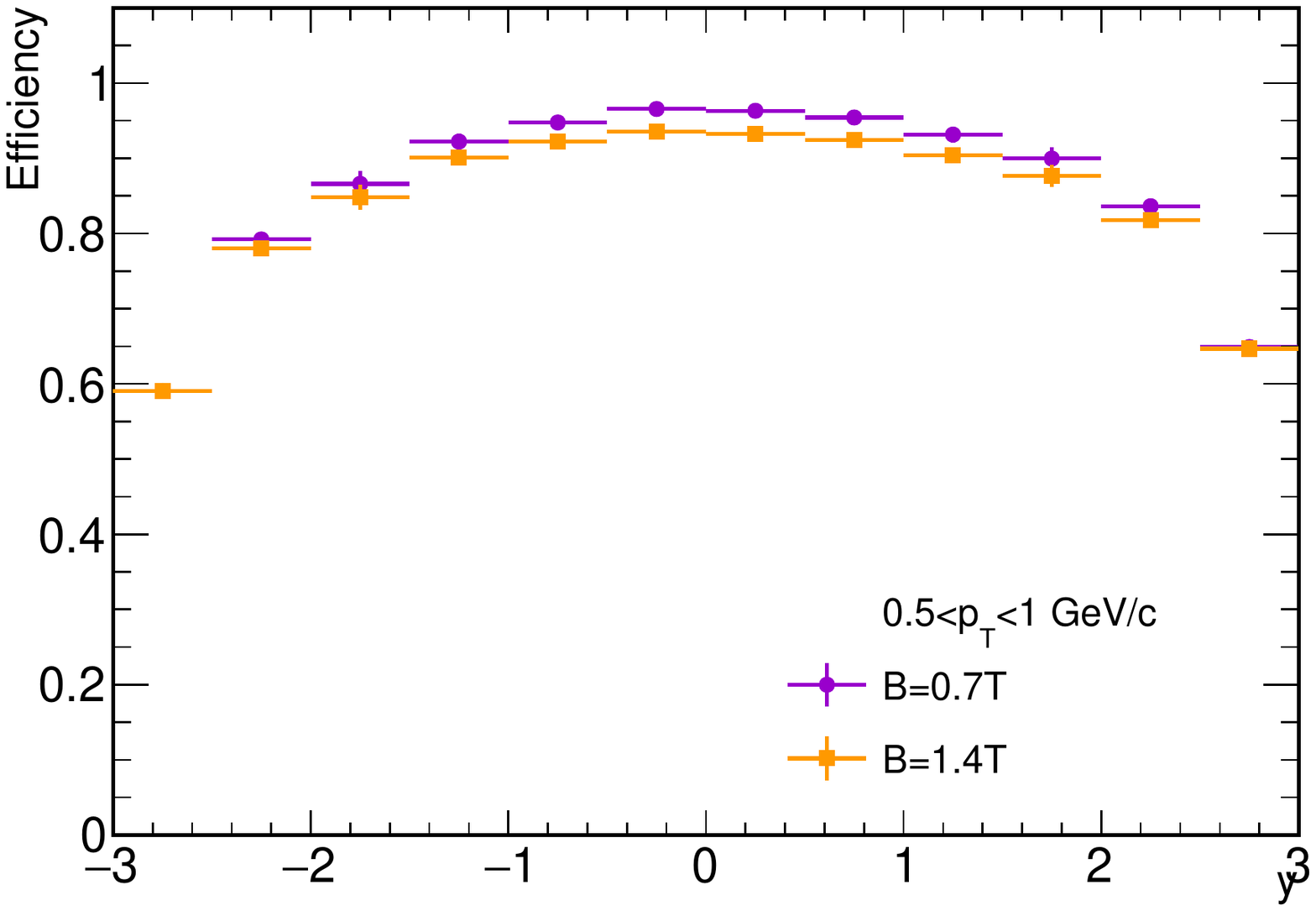}
        \includegraphics[width=0.37\linewidth]{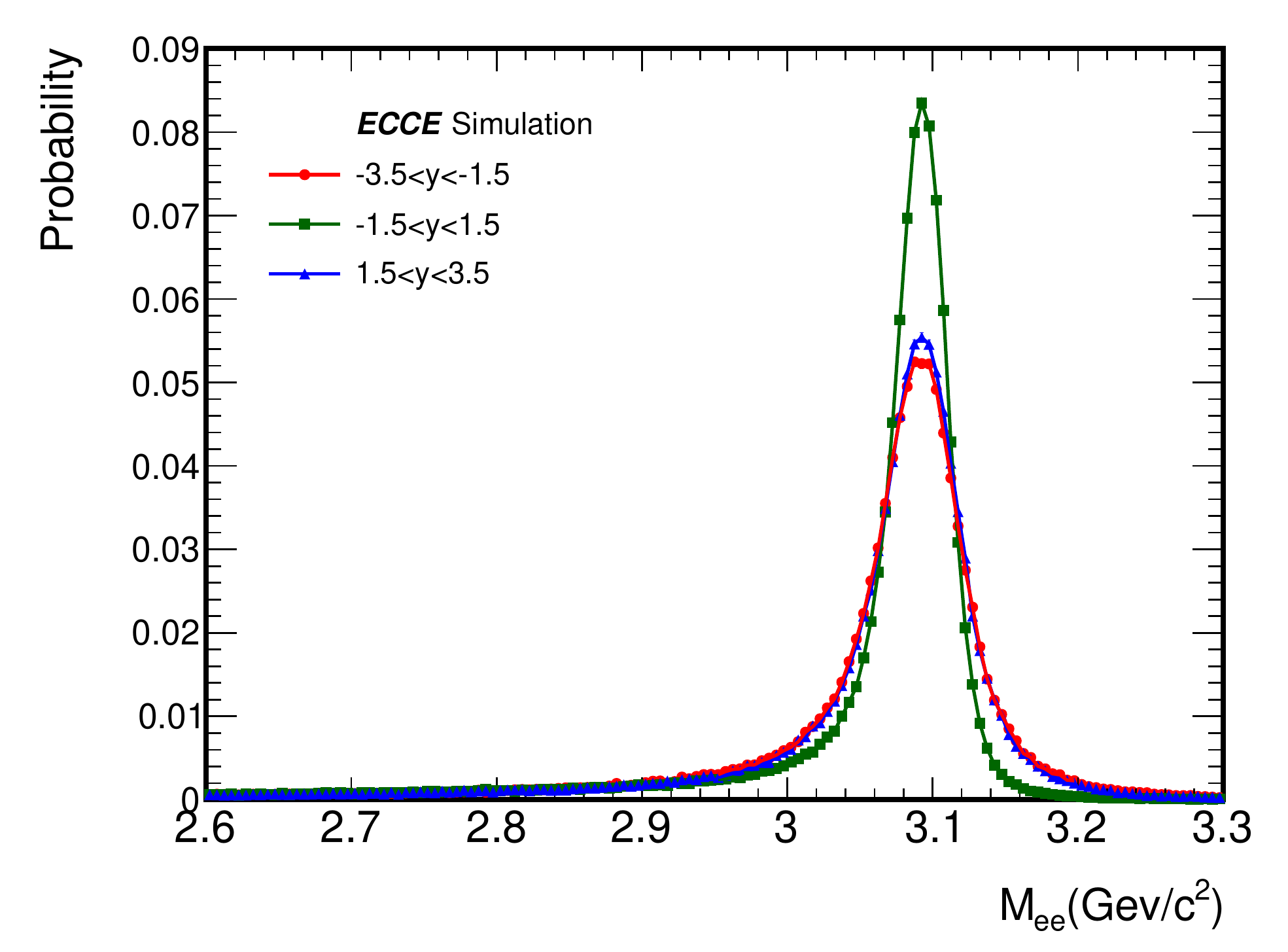}
    \caption{Magnetic strength effect on efficiency and bremsstrahlung energy loss effect on J/$\psi$ reconstruction.}
    \label{fig:tail}
\end{figure*}

 The kinematic distribution of J/$\psi$ for exclusive photoproduction is initialized by the theoretical calculation from eSTARLight. With this as input, we can obtain J/$\psi$ reconstruction efficiency from the Fun4All package with ECCE detector setup seen in Fig.~\ref{fig:singlejpsi}. The efficiency of J/$\psi$ is almost independent of the rapidity and transverse momentum except for the edge area at large forward and backward rapidity. We also study the effect of magnetic field strength and bremsstrahlung energy loss of electron on J/$\psi$ detection, shown as Fig.~\ref{fig:tail}. At very low $p_T$ ($0.5<p_T<1.0$ GeV/c), the improvement of the acceptance of the lower magnetic field strength accounts for the higher efficiency. While at larger $p_T$ ($1.0<p_T<2.0$ GeV/c), there is no significant difference between efficiencies of 0.7 Tesla and 1.4 Tesla. The bremsstrahlung energy loss has already been put in tracking performance in the "Prop.7" concept detector, which constitutes the tail in the reconstructed mass distribution depicted as the right panel in Fig.~\ref{fig:tail}. We scale the mass distribution to unity for the convenience of comparison, and the efficiencies of several mass window cuts are detailed in Table.~\ref{tab:tabel}. As expected, the tail effect is more significant for the J/$\psi$ at forward and backward rapidities (larger momentum of decayed electrons than that at central rapidity). With a proper mass cut window, the efficiency loss is minimal, implying that the effect of bremsstrahlung on J/$\psi$ reconstruction with ECCE setup is not significant.   
 
\begin{figure}
    \centering
        \includegraphics[width=0.8\linewidth]{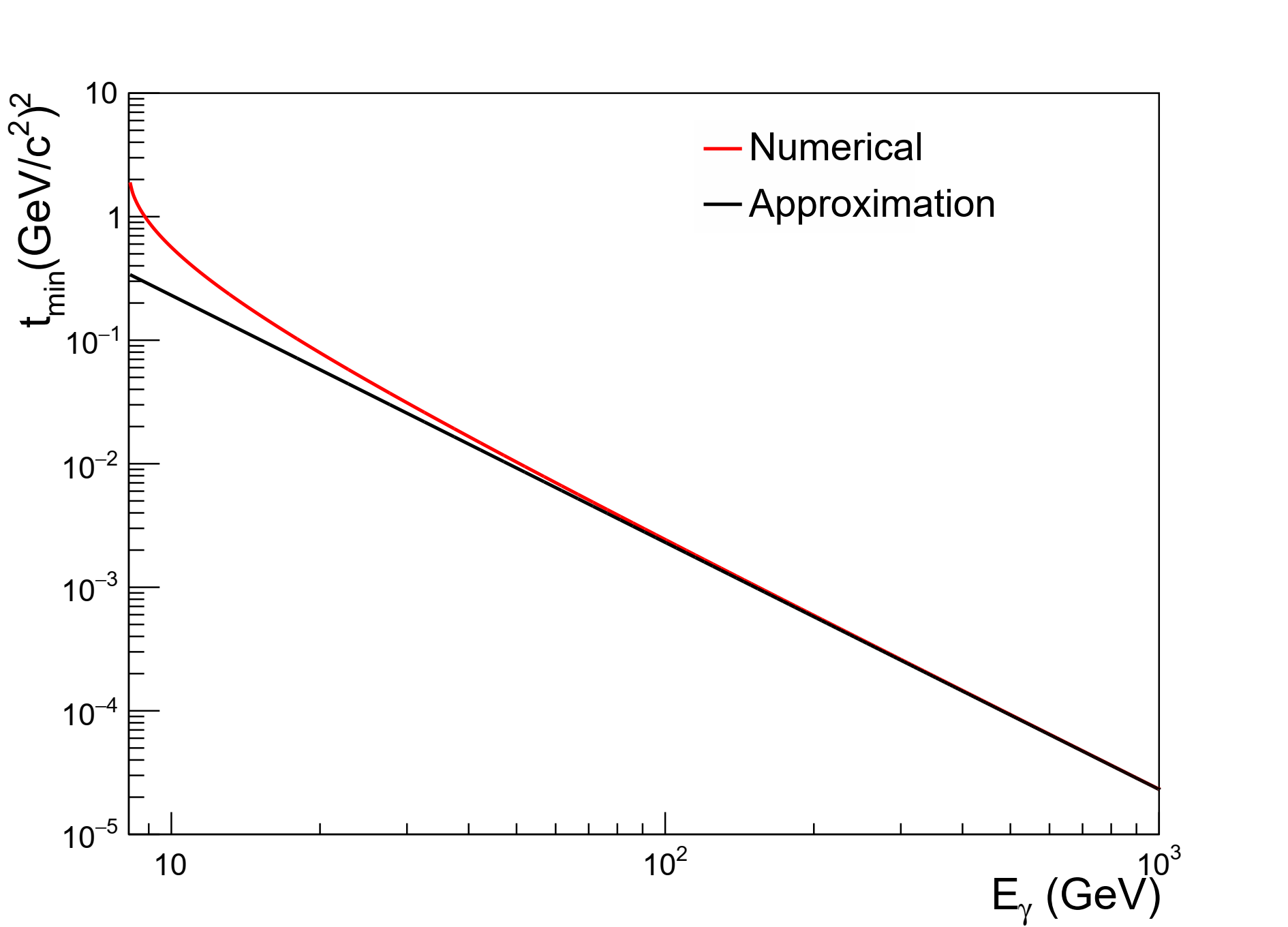}
        \includegraphics[width=0.8\linewidth]{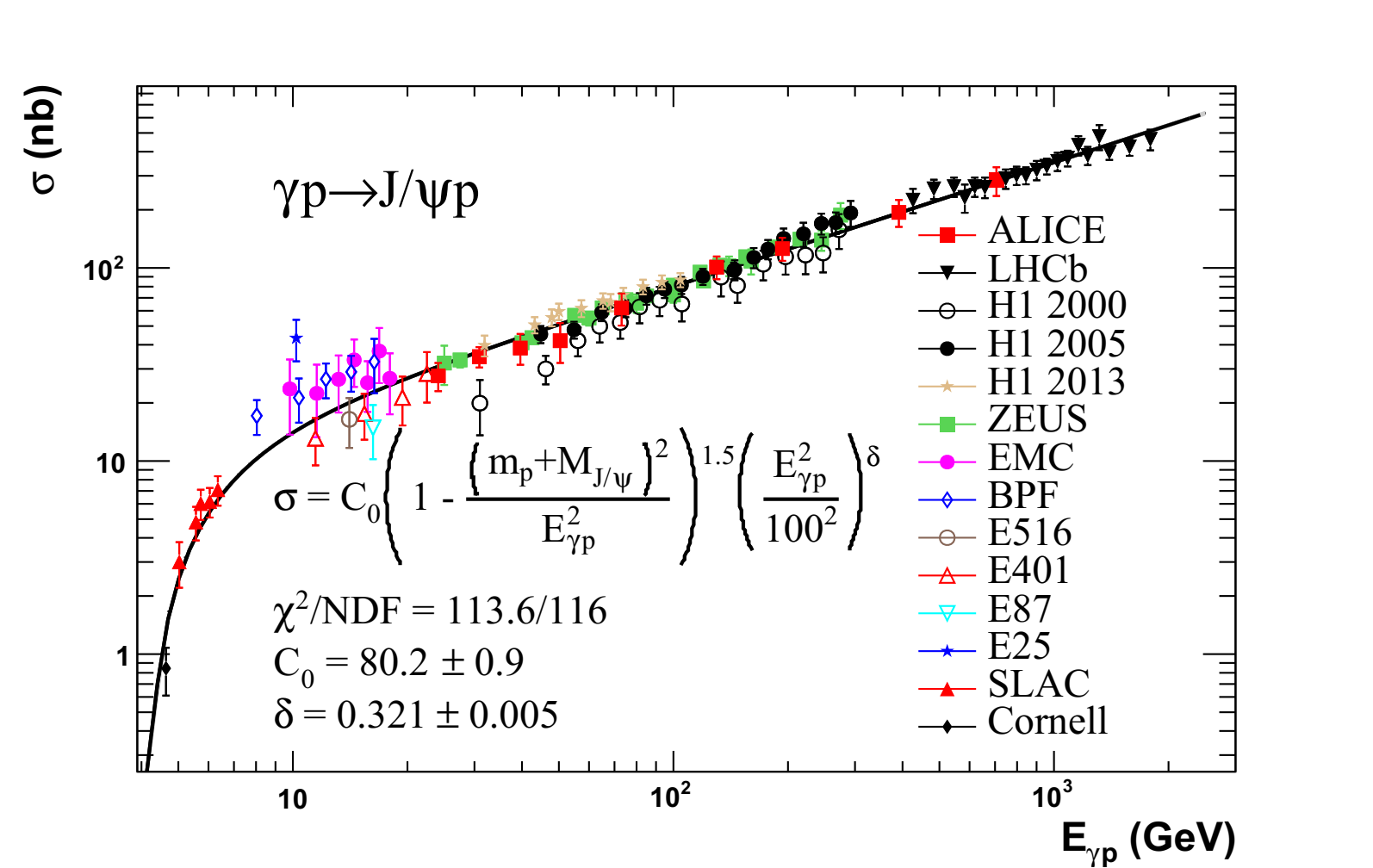}
    \caption{Upper Panel: The minimum momentum transfer as a function of incident photon energy in the rest frame of the nuclear beam (target frame). Lower Panel: The world-wide measurements of  $\sigma(\gamma p \rightarrow V p)$. }
    \label{fig:improve}
\end{figure}

\begin{table}[htbp]
    \centering
    \caption{Efficiency of mass window cut for J/$\psi$ reconstruction}
    \begin{tabular}{c|c|c|c}
         \hline
         mass  & \multirow{2}{*}{$-3.5<y<-1.5$}& \multirow{2}{*}{$-1.5<y<1.5$} & \multirow{2}{*}{$1.5<y<3.5$}\\
         window& & &\\
         (GeV/$c^2$)&  &  &\\
         \hline
         2.8-3.2 & 0.931    & 0.943 & 0.934\\
         2.9-3.2 & 0.903	& 0.917 & 0.907\\
         3.0-3.2 & 0.835	& 0.866 & 0.843\\
         \hline
    \end{tabular}
    \label{tab:tabel}
\end{table}

\begin{figure*}[!h]
    \centering
        \includegraphics[width=0.37\linewidth]{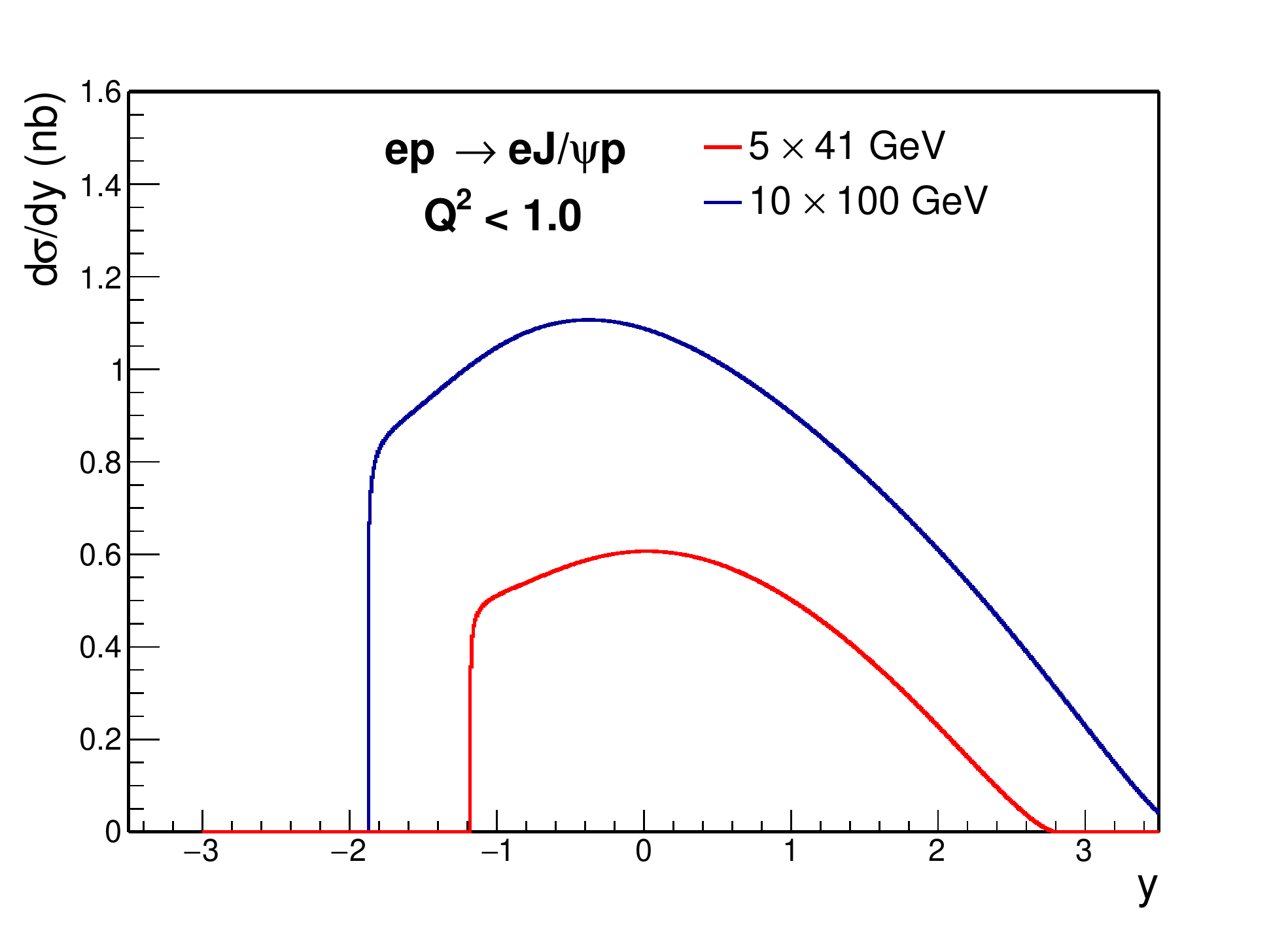}
        \includegraphics[width=0.37\linewidth]{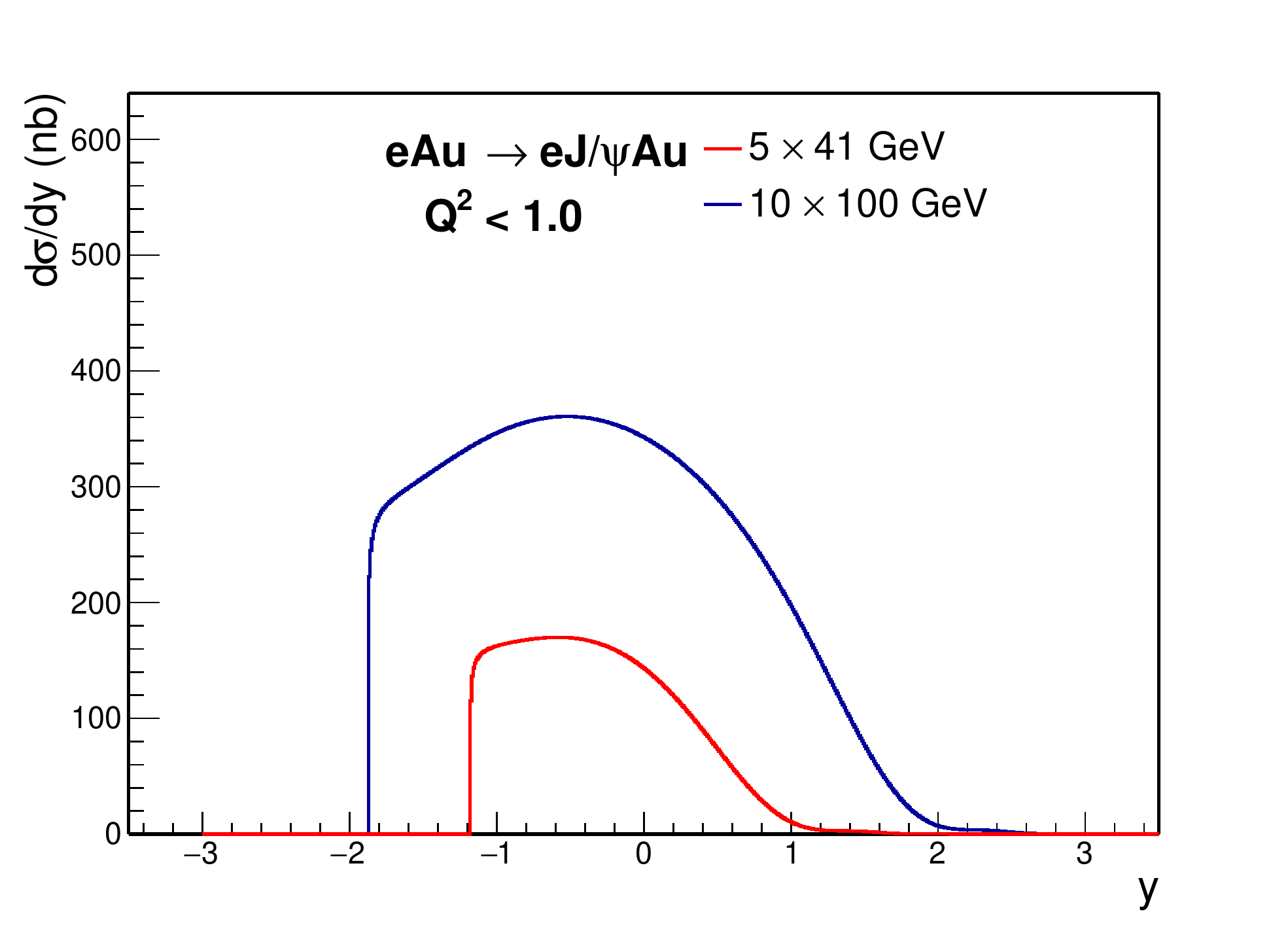}
    \caption{Rapidity dependence of differential cross section of exclusive J/$\psi$ photoproduction for $Q^2<1 GeV^2$.}
    \label{fig:y_distribution_1}
\end{figure*}

\begin{figure}
    \centering
    \includegraphics[width=0.8\linewidth]{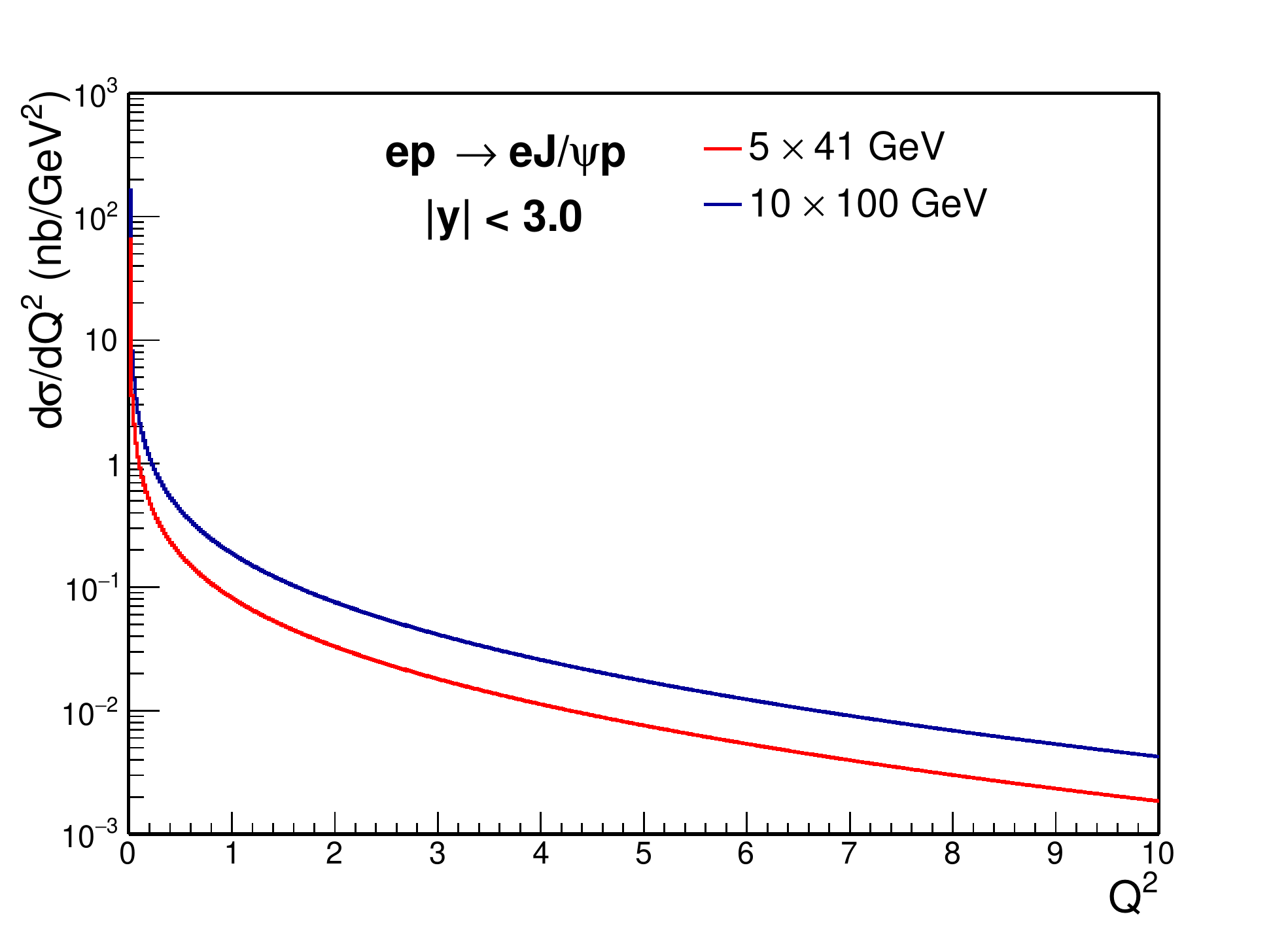}
    \caption{The $Q^2$ dependence of differential cross section of exclusive J/$\psi$ photoproduction in e+p collision.}
    \label{fig:Q_distribution_1}
\end{figure}

\begin{figure*}[!h]
    \centering
        \includegraphics[width=0.37\linewidth]{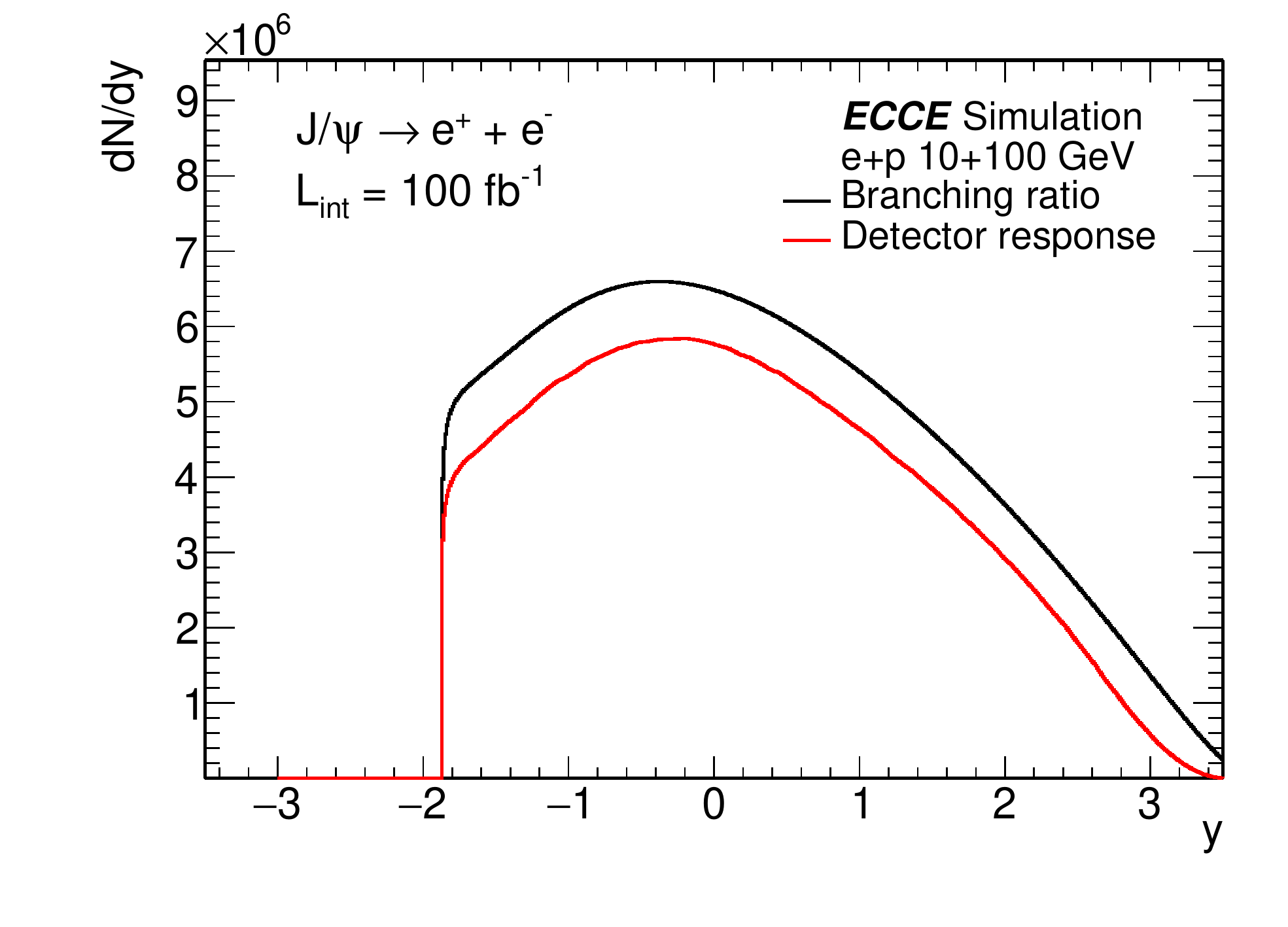}
        \includegraphics[width=0.37\linewidth]{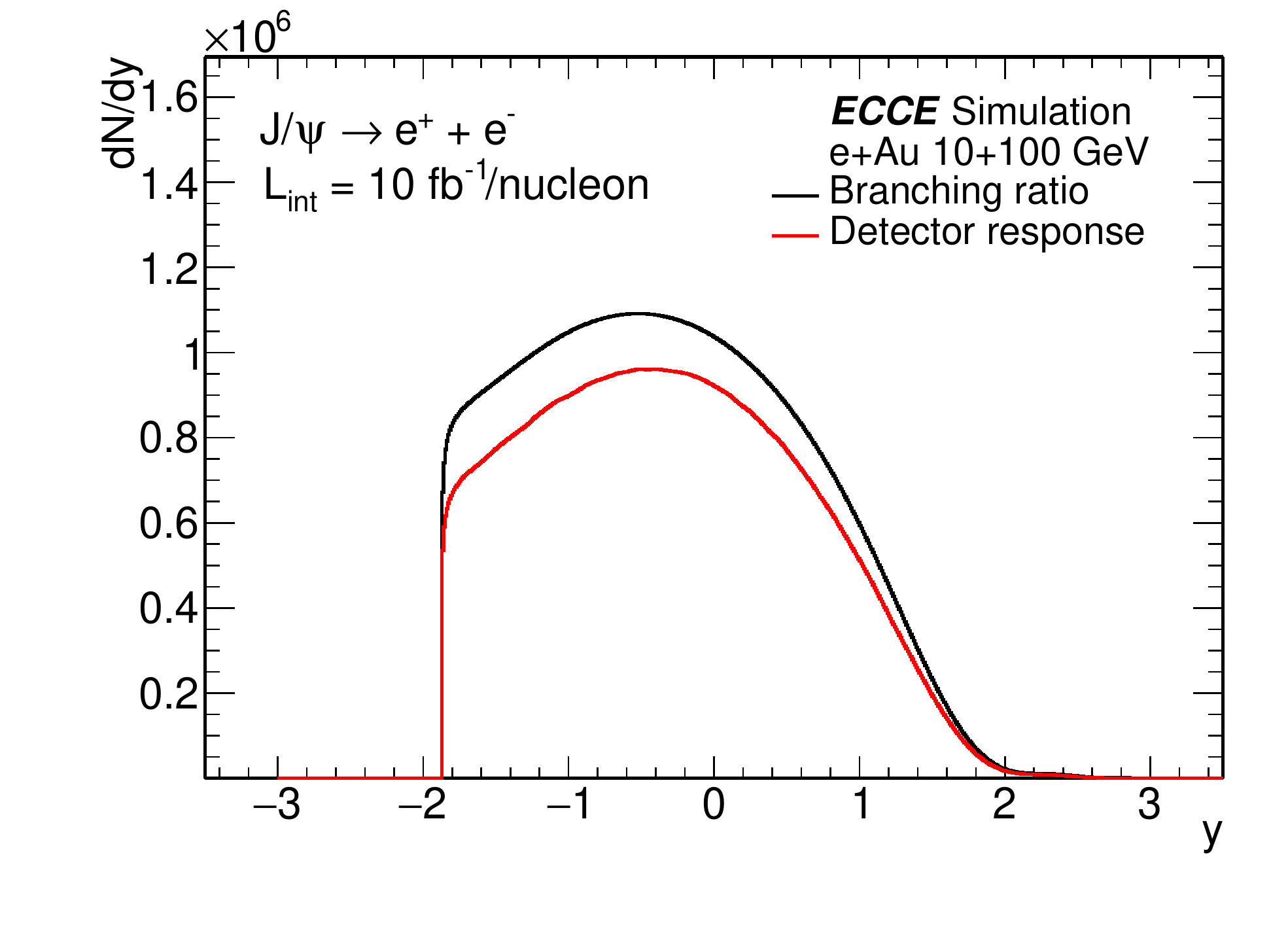}
    \caption{Rapidity dependence statistics of coherent exclusive production of J/$\psi$ in e+p and e+A collisions for 10$\times$100 GeV. Left Panel: e+p collision. Right Panel: e+Au collision}
    \label{fig:y_distribution_2}
\end{figure*}

\section{Theoretical Setup for Projection}
\label{theoretical:setup}
\indent This section presents the theoretical framework of exclusive J/$\psi$ photoproduction in e+p and e+A collisions, which is employed in the simulation. The cross section of exclusive vector meson photoproduction $\sigma(e A \rightarrow e A V)$ is calculated as an integration of photon flux induced by the electron beam and the collision of the virtual photon on the target nucleus. The cross section of exclusive vector meson photoproduction $\sigma(e A \rightarrow e A V)$ is derived by integrating the photon flux caused by the electron beam and the virtual photon collision on the target nucleus, as illustrated in Eq.(\ref{1}): 

\begin{equation}
\label{1}
\sigma(e A \rightarrow e A V)=\int \frac{d W}{W} \int d k \int d Q^{2} \frac{d^{2} N_{\gamma}}{d k d Q^{2}} \sigma_{\gamma^{*} A \rightarrow V A}\left(W, Q^{2}\right),
\end{equation}
where the photon flux can be written as:

\begin{equation}
\label{2}
\frac{d^{2} N_{\gamma}}{d k d Q^{2}}=\frac{\alpha}{\pi k Q^{2}}\left[1-\frac{k}{E e}+\frac{k^{2}}{2 E_{e}^{2}}-\left(1-\frac{k}{E e}\right)\left|\frac{Q_{\min }^{2}}{Q^{2}}\right|\right].
\end{equation}
The cross section of virtual photon collision on the nucleus can be related to the production cross section with real photon:

\begin{equation}
\label{3}
\begin{aligned}
\sigma_{\gamma^{*} A \rightarrow V A}\left(W, Q^{2}\right)=&f\left(M_{V}\right) \sigma\left(W, Q^{2}=0\right)\left(\frac{M_{V}^{2}}{M_{V}^{2}+Q^{2}}\right)^{n} \\&n=c_{1}+c_{2}\left(Q^{2}+M_{V}^{2}\right),
\end{aligned}
\end{equation}
where $c_1$ and $c_2$ are parameters determined by the HERA measurements. $f\left(M_{V}\right)$ is the Breit-Wigner distribution of the vector meson. And the cross section at $Q^2=0$ can be calculated by the integration of the forward scattering cross section and the square of the nucleus form factor, revealed as Eq.(\ref{4}):

\begin{equation}
\label{4}
\sigma\left(W, Q^{2}=0\right)=\int_{t_{\min }}^{\infty}\left.d t \frac{d \sigma(\gamma A \rightarrow V A)}{d t}\right|_{t=0} |F(t)|^{2},
\end{equation}
where $\frac{d \sigma(\gamma A \rightarrow V A)}{d t}|_{t=0}$ can be determined by $\frac{d \sigma(\gamma p \rightarrow V p)}{d t}|_{t=0}$ via Glauber approach. The cross section of $\gamma p \rightarrow V p$ can be parameterized  using the world-wide measurements~\cite{Chinesec}. The framework is almost the same as eSTARLight~\cite{PhysRevC.99.015203}\cite{PhysRevC.97.044910}, except for two minor improvements. In eSTARLight, the minimum momentum transfer $t_{min}$ is approximated as $t_{min} = ((M_v)^2 / 2 k)^2$. One can get the minimum of t when the transverse momentum of the produced vector meson is equal to zero. Then true $t_{min}$ can be obtained from energy-momentum conservation of the $\gamma p\rightarrow V p$ process in the target frame as Eq.(\ref{22}):

\begin{equation}
\label{22}
\begin{aligned}
E_{\gamma}+m_{p}=\sqrt{M_{v}^{2}+\left(E_{\gamma}-P_{z}^{\prime 2}\right)}+\sqrt{m_{p}^{2}+P_{z}^{\prime 2}},
\end{aligned}
\end{equation}

\begin{equation}
\label{23}
\begin{aligned}
t=\left(P^{\prime}-P\right)^{2}=\left(\sqrt{m_{p}^{2}+P_{z}^{\prime 2}}-m_{p}\right)^{2}-P_{z}^{\prime 2},
\end{aligned}
\end{equation}

where $P_{z}^{\prime 2}$ is the longitudinal momentum of the final state proton. The photon energy dependence of $t_{min}$ can be found in the upper panel of Fig.~\ref{fig:improve}. The approximation in eSTARLight is proper at high photon energy. However, it would underestimate the magnitude at low values of photon energy, as is the case for our projection at ECCE. Furthermore, in eSTARLight, the parametrization of the $\gamma p \rightarrow V p$ cross section is only based on high-energy HERA data. The behavior of energy dependency is notably different between the high and low energy ranges, as demonstrated in the lower panel of Fig.~\ref{fig:improve}, which would skew the computations at EIC. With these two improvements, the calculated results of rapidity distribution for exclusive J/$\psi$ photoproduction in e+p and e+A collisions for 5$\times$41 and 10$\times$100 GeV collision energies are shown in Fig.~\ref{fig:y_distribution_1}. $Q^2$ dependence of e+p collision for 5$\times$41 and 10$\times$100 GeV with a rapidity range from $-3$ to 3 is illustrated in Fig.~\ref{fig:Q_distribution_1}.     \newline



\begin{figure*}[!h]
    \centering
        \includegraphics[width=0.37\linewidth]{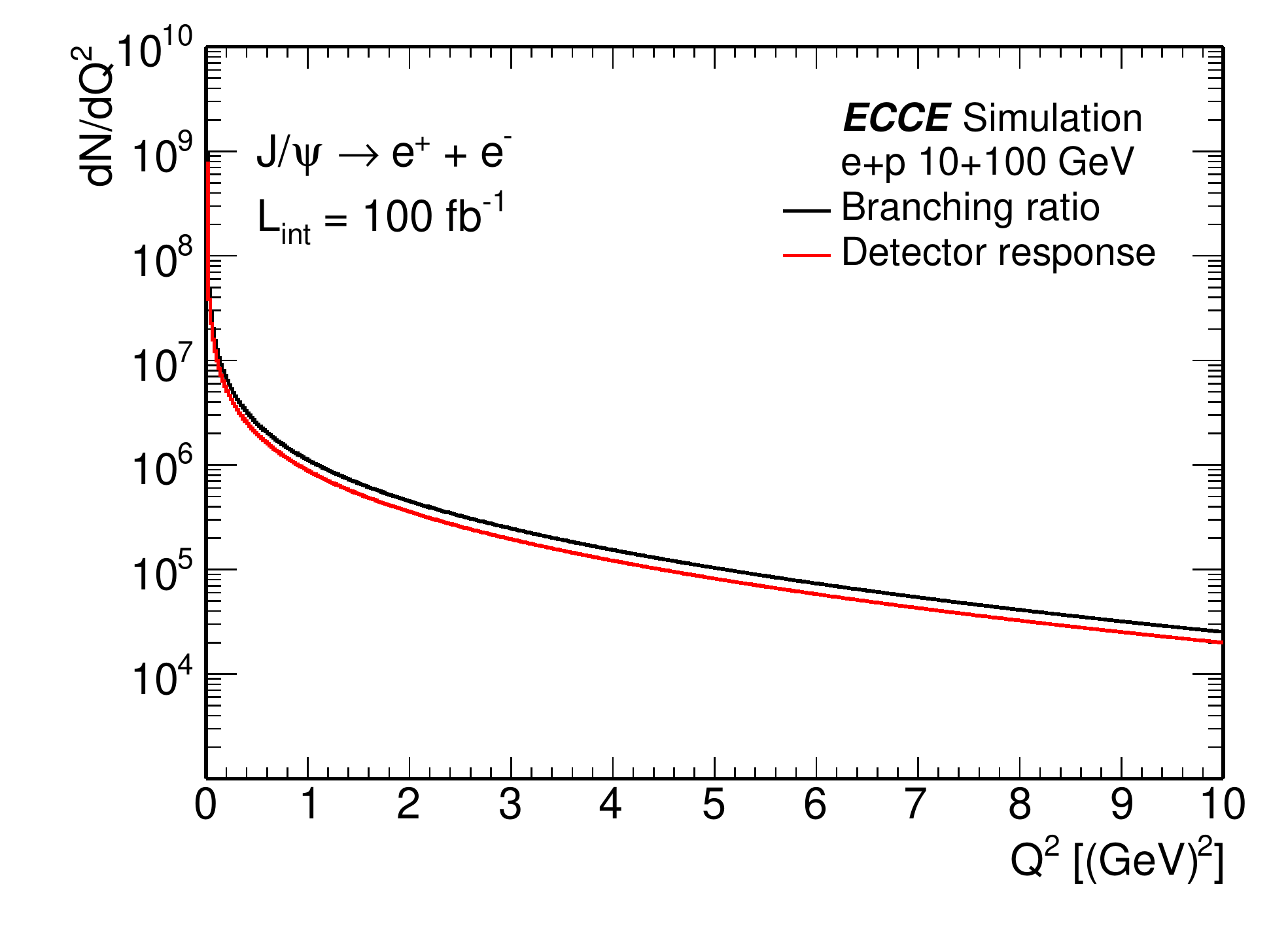}
        \includegraphics[width=0.37\linewidth]{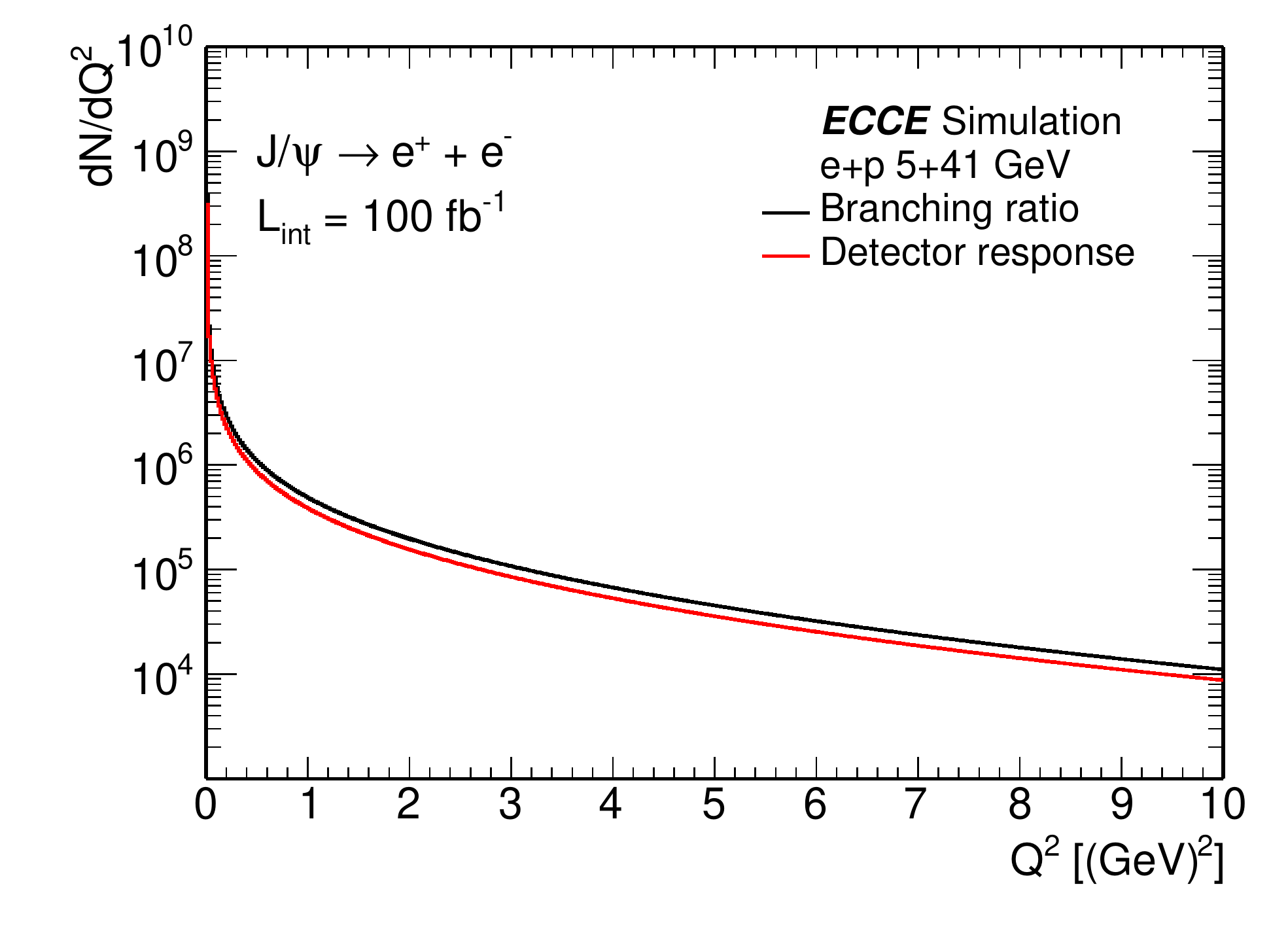}
    \caption{The $Q^2$ dependence of J/$\psi$ photoproduction in e+p. Left Panel: 10$\times$100 GeV. Right Panel: 5$\times$41 GeV.}
    \label{fig:Q_distribution_2}
\end{figure*}

The raw counts per unit rapidity are shown in Fig.~\ref{fig:y_distribution_2} for e+p and e+Au collisions for 10$\times$100 GeV. For the projection results in the following section, we assume the integrated luminosity collected by ECCE is $100 fb^{-1}$ for e+p collisions and $10 fb^{-1} /A$ for e+Au collisions, where A is the mass number of Au. The figure shows that millions of J/$\psi$s would be observed with the designed ECCE setup, which provides us with plenty of physics opportunities. And $Q^2$ dependence of the statistics of e+p collision in 5$\times$41 and 10$\times$100 GeV are shown in Fig.~\ref{fig:Q_distribution_2}. As we can see, most events locate in the low $Q^2$ region, especially for $Q^2<1$ $GeV^2$.

\section{Physics Opportunities with Exclusive J/$\psi$ Photoproduction at ECCE}

\subsection {Probe the nuclear gluon PDF}
\label{g_dep}

The gluon parton distribution functions (PDFs) in the proton and nucleus have large uncertainties because gluons do not carry any electric charge and can not be directly determined by the DIS measurements. As mentioned in the introduction, the exclusive J/$\psi$ photoproduction occurs via Pomeron exchange. Due to the gluonic nature of Pomeron, this process is directly sensitive to the gluon PDF. According to the calculation of perturbative QCD, the forward scattering cross section is proportional to the square of the gluon density distribution, shown in the following~\cite{2013}\cite{2014}:

\begin{equation}
\label{5}
\left.\frac{d \sigma(\gamma A \rightarrow V A)}{d t}\right|_{t=0}=\frac{\alpha_{s}^{2} \Gamma_{e e}}{3 \alpha M_{V}^{\xi}} 16 \pi^{3}\left\lfloor x g_{A} (x, \mu^{2})\right\rfloor^{2},
\end{equation}

where $\Gamma_{e e}$ is the width of the electronic decay of J/$\psi$, $g_{A} (x, \mu^{2})$ is the gluon density and the momentum fraction $x$ can be determined by the rapidity of J/$\psi$:

\begin{equation}
\label{6}
x=\frac{M_{V} e^y}{2 E_N},
\end{equation}

where $E_N$ is the energy of nuclear beam per nucleon. Eq.(\ref{5}) is derived from leading order (LO) pQCD calculation in the non-relativistic approximation~\cite{2013}, which indicates that the transverse momenta of c quarks in J/$\psi$ are negligible. In that case, it is prescribed that $\mu^{2}=M_{V}^{2} / 4$. \newline

The nuclear gluon shadowing can be model-independently quantified by $R_{g}$:
\begin{equation}
\label{20}
R_{g}=\sqrt{\frac{\left.\frac{d \sigma(\gamma A \rightarrow V A)}{d t}\right|_{t=0}}{\left.\frac{d \sigma(\gamma p \rightarrow V p)}{d t}\right|_{t=0}}}.
\end{equation}

As shown in Eq.(\ref{5}), if we make the forward scattering amplitude ratio between e+p and e+Au collisions, the shadowing factor $R_g$ of gluon can be directly extracted. So measurements of J/$\psi$ photoproduction can provide direct access to $g_A ( x, \mu^2)$.

\begin{figure}
    \centering
    \includegraphics[width=0.8\linewidth]{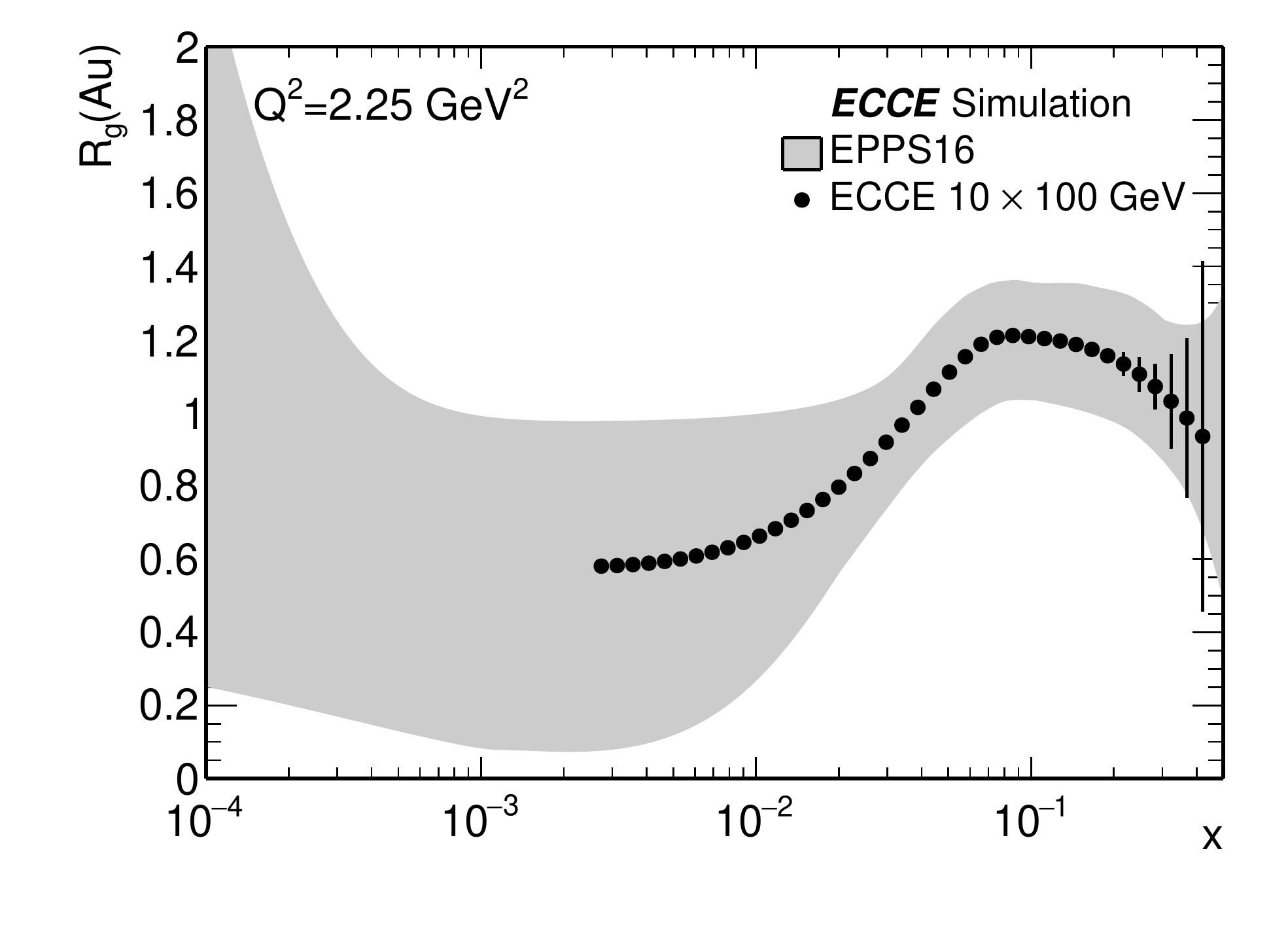}
    \caption{Gluon nuclear shadowing factor as function of momentum fraction x.}
    \label{fig:gshadow}
\end{figure}

\indent Elastic J/$\psi$ photoproduction processes are simulated in 10$\times$100 (GeV) e+p and e+Au collisions with the framework described in the above sections. From the simulation, we extracted the $d^2 \sigma/dtdy$ of J/$\psi$ at t =0 for both e+p and e+Au collisions to make projection on $R_g$. The uncertainty of the projection only includes the statistical error. At a given $x$, we can transfer it to the corresponding $y$ value via Eq.(\ref{6}) and get the statistics with the detector response. Then we fit the simulated t distribution with the predicted statistics and get the fit error of $d \sigma / dt$ for e+p and e+Au collisions at $t=0$. The statistical error of projection can be extracted by the error propagation approach via $R_g$ equation in Eq.(\ref{6}). As shown in Fig.\ref{fig:gshadow}, the measurement of exclusive J/$\psi$ production has a wide x coverage down to 2$\times 10^{-3}$ for beam configuration 10$\times$100 GeV. In the low x region, the EPPS16~\cite{2017} PDF set has a large uncertainty band, while the projected statistical error for ECCE is negligible. This shows that the precision exclusive J/$\psi$ measurements at the EIC will significantly reduce the uncertainty of the nuclear gluon PDF at low values of x (x$<10^{-2}$).

\subsection{Probe the Gluon Spatial Distribution}

\label{t_dep}
The Pomeron is the exchange object for the diffractive process, and diffraction is generally sensitive to spatial distribution. The momentum transfer from the target in the exclusive J/$\psi$ photoproduction is sensitive to the production site, which provides us with a powerful tool to infer the spatial distribution of gluon in both proton and nucleus. 

\begin{figure}
    \centering
    \includegraphics[width=0.8\linewidth]{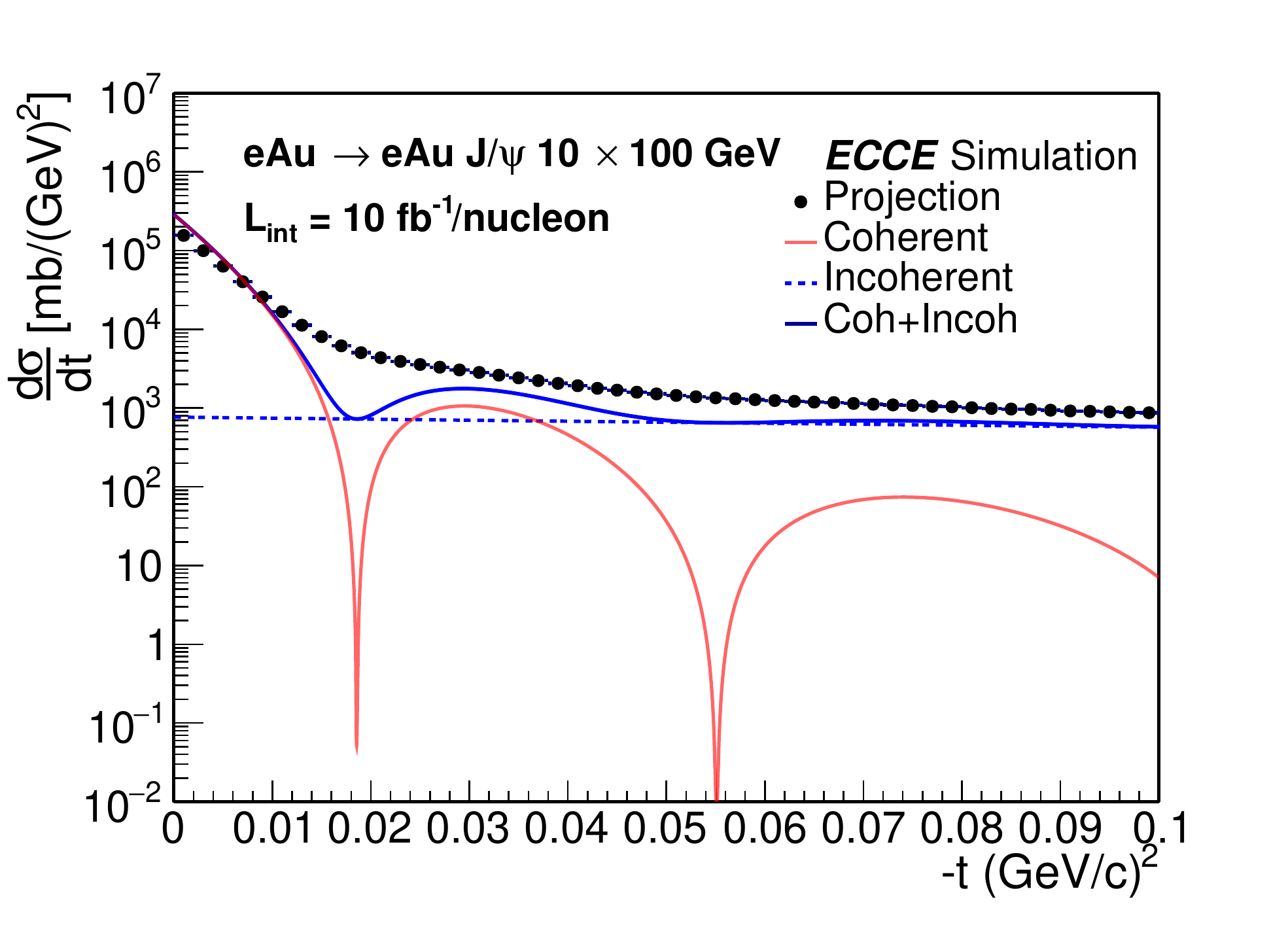}
    \caption{t dependence of exclusive J/$\psi$ production in e+Au collision. Wood-Saxon distribution is used as input.}
    \label{fig:t_distribution}
\end{figure}

In the simulation, the Woods-Saxon distribution is used as input of the gluon source distribution F(b). We made a projection of t distribution for the processes in 10 $\times$ 100 (GeV) e+Au collisions with both coherent and incoherent J/$\psi$ photoproduction. The results are shown in Fig.\ref{fig:t_distribution}, the red and blue curves are the coherent and incoherent contributions from calculations, respectively. The solid data points are the projected results from simulation, in which the statistical uncertainty is negligible. However, it should be noticed that according to the state of the art theoretical calculation of elastic scattering, we will not get this minimum at these dips~\cite{PhysRevC.78.044332}\cite{Liu_2019}. So the model here is a simplified and ideal one, and the purpose is to show the momentum resolution impact on this t dependence. Due to the momentum smearing from tracking system, the slope of the distribution is slightly different from that of theoretical calculations, and the diffraction dips are fold out. This suggests that the detector response should be precisely determined to extract the gluon spatial distribution. \newline

\begin{figure*}[!h]
    \centering
    \includegraphics[width=0.37\linewidth]{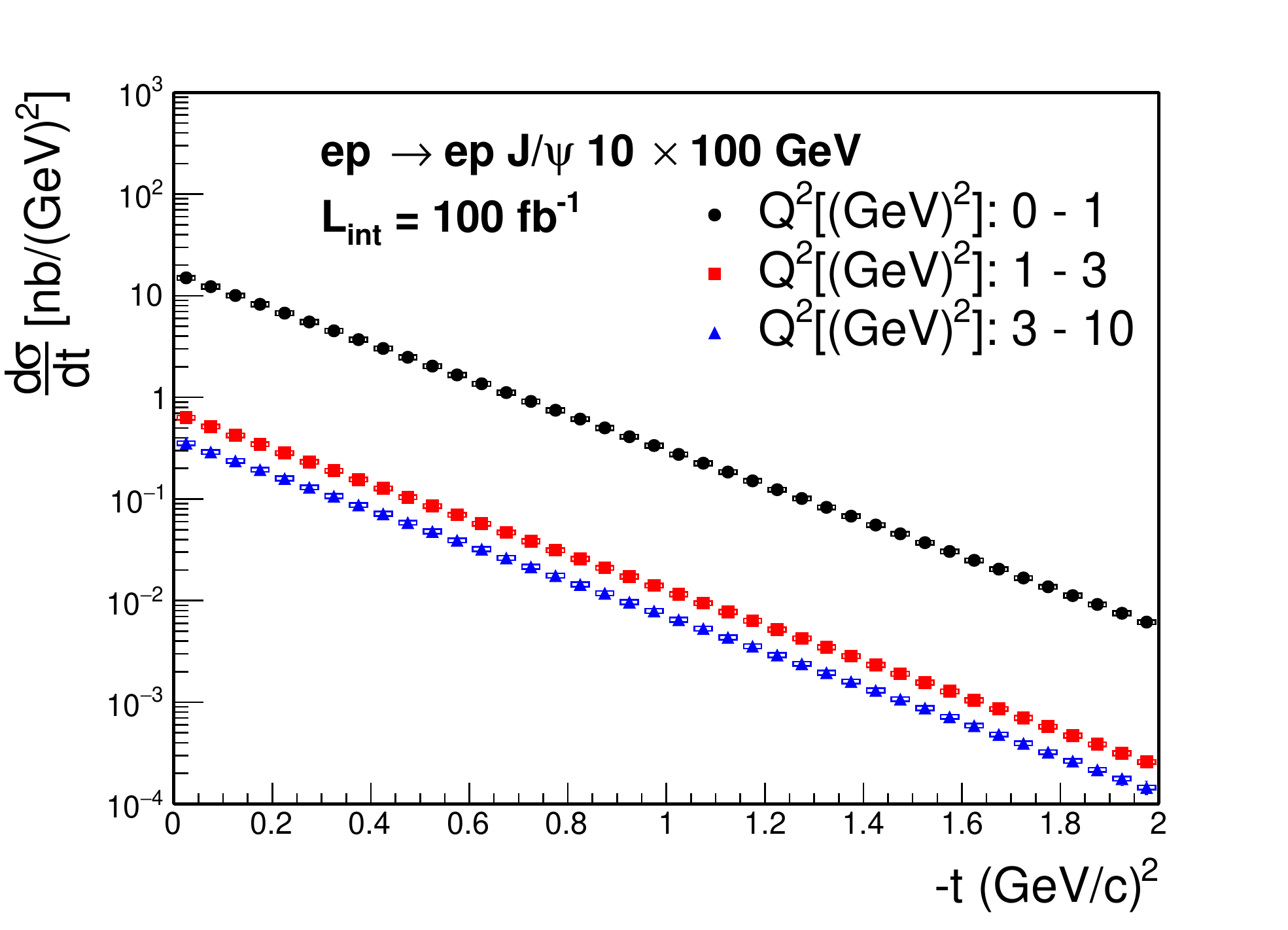}
    \includegraphics[width=0.37\linewidth]{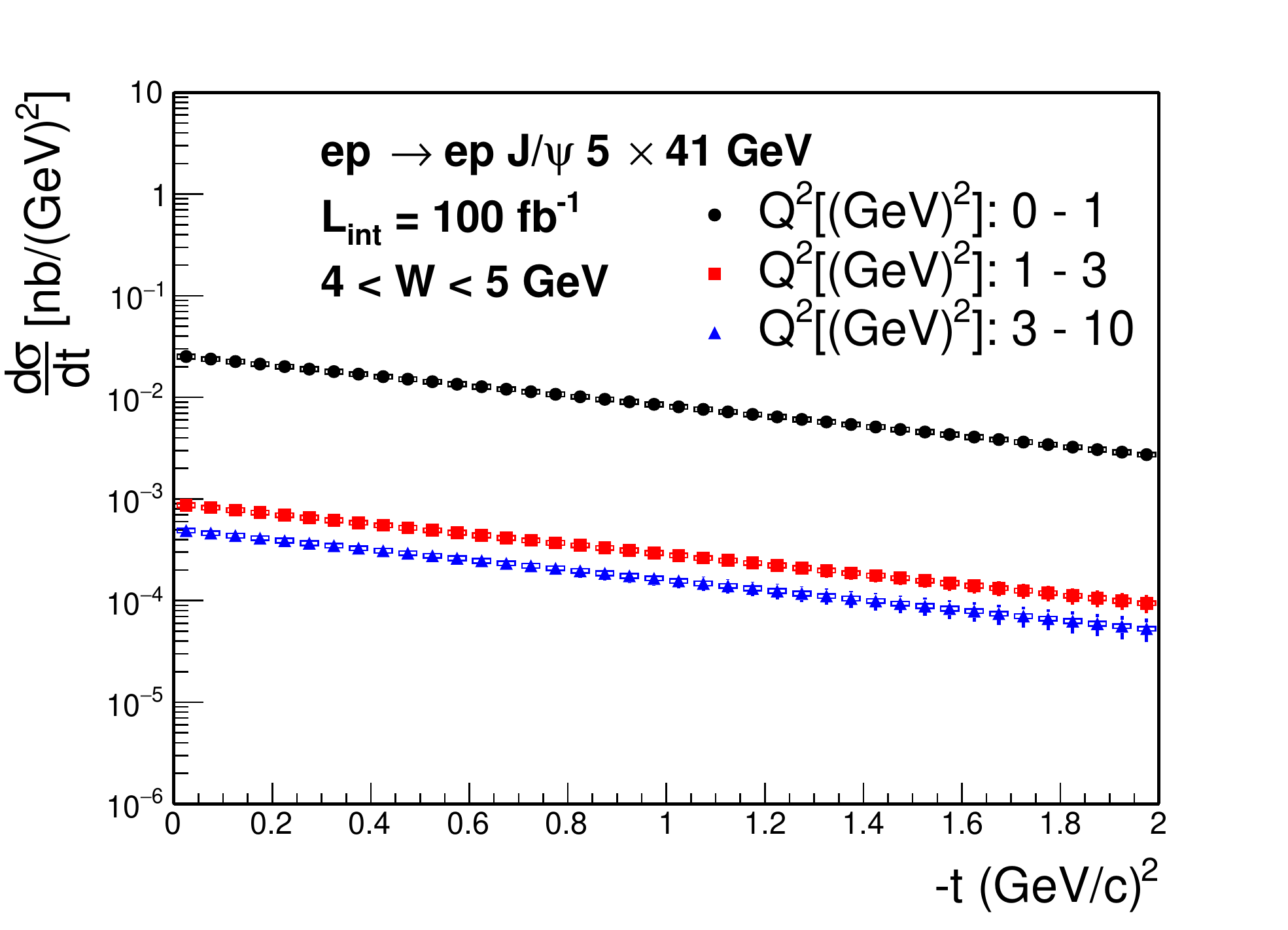}
    \caption{t dependence of exclusive J/$\psi$ production in e+p collision in several $Q^2$ intervals. Left Panel: 10$\times$100. Right Panel: 5$\times$41}
    \label{fig:t_distribution_2}
\end{figure*}

The projected t distributions in 5 $\times$ 41 and 10 $\times$ 100 (GeV) e+p collisions are illustrated in Fig.~\ref{fig:t_distribution_2} with the systematical and the statistical errors for different $Q^2$ regions (0-1, 1-3 and 3-10 $GeV^2$). The statistical error is determined with a similar approach given in Sec.\ref{g_dep}. The systematical uncertainty is determined by the maximum deviation between the input and reconstructed t distribution. The statistical and systematical uncertainties are added in quadrature.

\subsection {The Near-Threshold Production Mechanism}
\label{thresh}

The elastic near-threshold J/$\psi$ production can provide new insight into multi-quark, gluonic, hidden-color correlations of hadronic and nuclear wave-functions in QCD. Moreover, the measurements of this process probe the $x \sim 1$ configuration in the target, and the spectator partons carry a vanishing fraction $x \sim 0$ of the target momentum. This implies that the production rate behaves near $x \rightarrow 1$ as $(1-x)^{2n_s}$, where $n_s$ is the number of spectators. Then two gluon and three gluon exchange contributions can be written as~\cite{2001}:

\begin{equation}
\label{7}
\frac{d \sigma}{d t}=\mathcal{N}_{2 g} v \frac{(1-x)^{2}}{R^{2} \mathcal{M}^{2}} F_{2 g}^{2}(t)\left(W_{\gamma p}^2-m_{p}^{2}\right)^{2},
\end{equation}
\begin{equation}
\label{8}
\frac{d \sigma}{d t}=\mathcal{N}_{3 g} v \frac{(1-x)^{0}}{R^{4} \mathcal{M}^{4}} F_{3 g}^{2}(t)\left(W_{\gamma p}^2-m_{p}^{2}\right)^{2},
\end{equation}
where R is the radius of proton, M is the mass of J/$\psi$, and $W_{\gamma p}$ is the center of mass energy of $\gamma$p.

\begin{figure}
    \centering
    \includegraphics[width=0.8\linewidth]{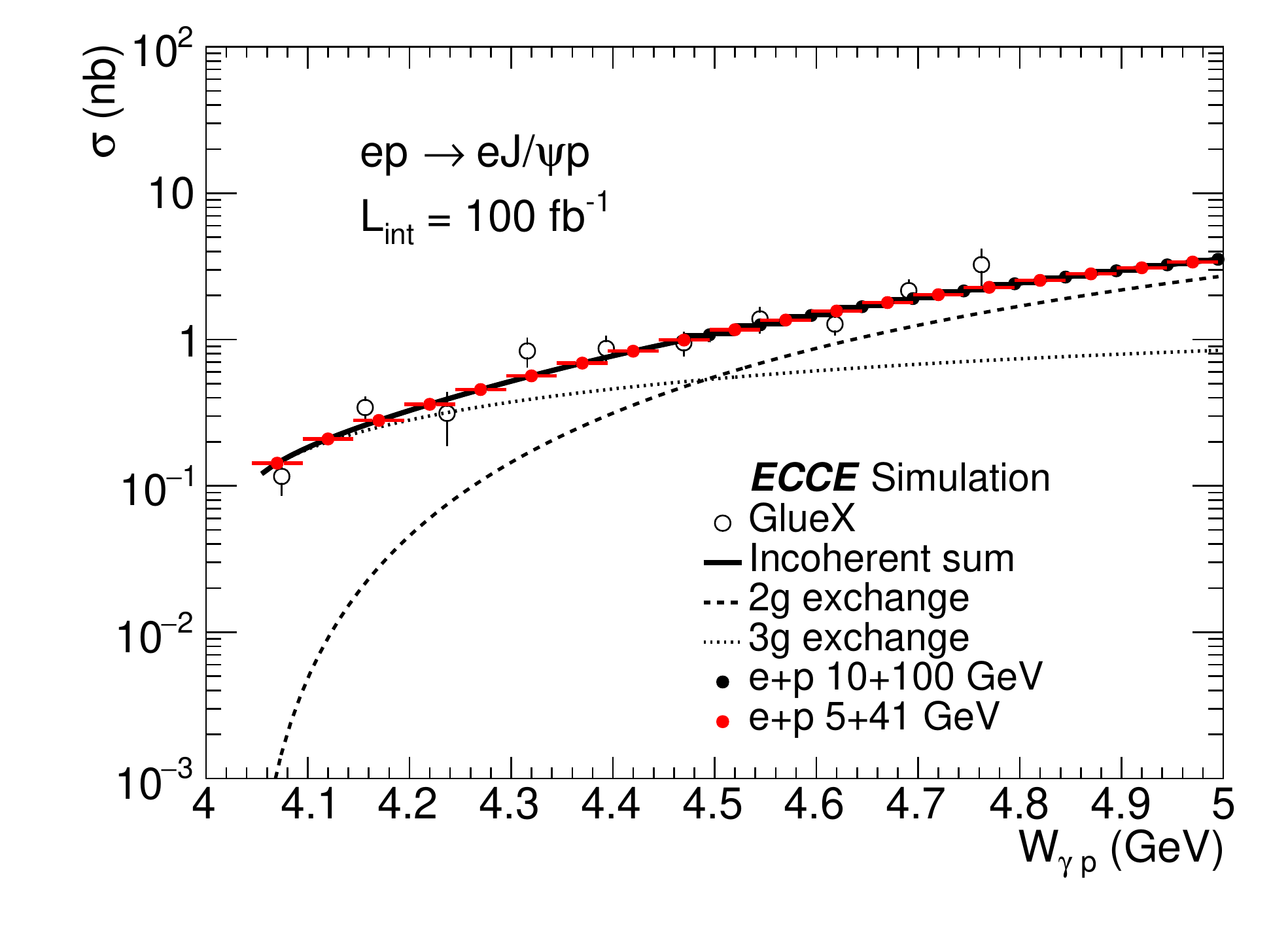}
    \caption{Projection of J/$\psi$ photoproduction cross section near threshold in 10$\times$100 GeV and 5$\times$41 GeV e+p collisions.}
    \label{fig:near}
\end{figure}

The projected results for near-threshold production for 10$\times$100 GeV and 5$\times$41 GeV e+p collisions are shown in Fig.~\ref{fig:near}. The GlueX results, two and three gluon exchange contributions are also shown for comparison. All the theoretical curves and projection results are normalized with the GlueX measurements. The error bars on GlueX measurements represent only statistical uncertainties. At low $W_{\gamma p} < 4.5 $ GeV region, the cross section is dominated by three gluon exchange process. At $W_{\gamma p} > 4.5  $ GeV , two gluon exchange process comes to take control. For 10$\times$100 GeV e+p collisions, the center-of-mass energy can only reach as low as 4.5 GeV due to the limited detector coverage. But for 5$\times$41 GeV e+p collisions, they cover the whole near-threshold range. The GlueX experiment at JLab has already shed light on the near-threshold production mechanism as a sum of two-gluon and three-gluon exchange and set limits on pentaquark production~\cite{new6}. Measurements of near-threshold with larger statistics and broader $W_{\gamma p}$ range at the EIC has the potential to impose more powerful constraints on the production mechanism, like charmed pentaquark $P_c$ production~\cite{new5,new4}. 

\subsection{Trace Anomaly and Proton Mass Decomposition}
\label{trace}

According to QCD theory, there are four terms of decomposition in nucleon mass as Eq.(\ref{9})~\cite{refId0}: quark energy $M_{q}$, gluon energy $M_{g}$, quark mass $M_{m}$ and the trace anomaly contribution $M_{a}$, and these terms are sensitive to the momentum fraction a carried by all quarks and the trace anomaly parameter b.

\begin{equation}
\begin{aligned}
\label{9}
&M_{q}=\frac{3}{4}\left(a-\frac{b}{1+\gamma_{m}}\right) M_{N},\\
&M_{g}=\frac{3}{4}(1-a) M_{N},\\
&M_{m}=\frac{4+\gamma_{m}}{4\left(1+\gamma_{m}\right)} b M_{N},\\
&M_{a}=\frac{1}{4}(1-b) M_{N},
\end{aligned}
\end{equation}

\begin{figure}
    \centering
    \includegraphics[width=0.8\linewidth]{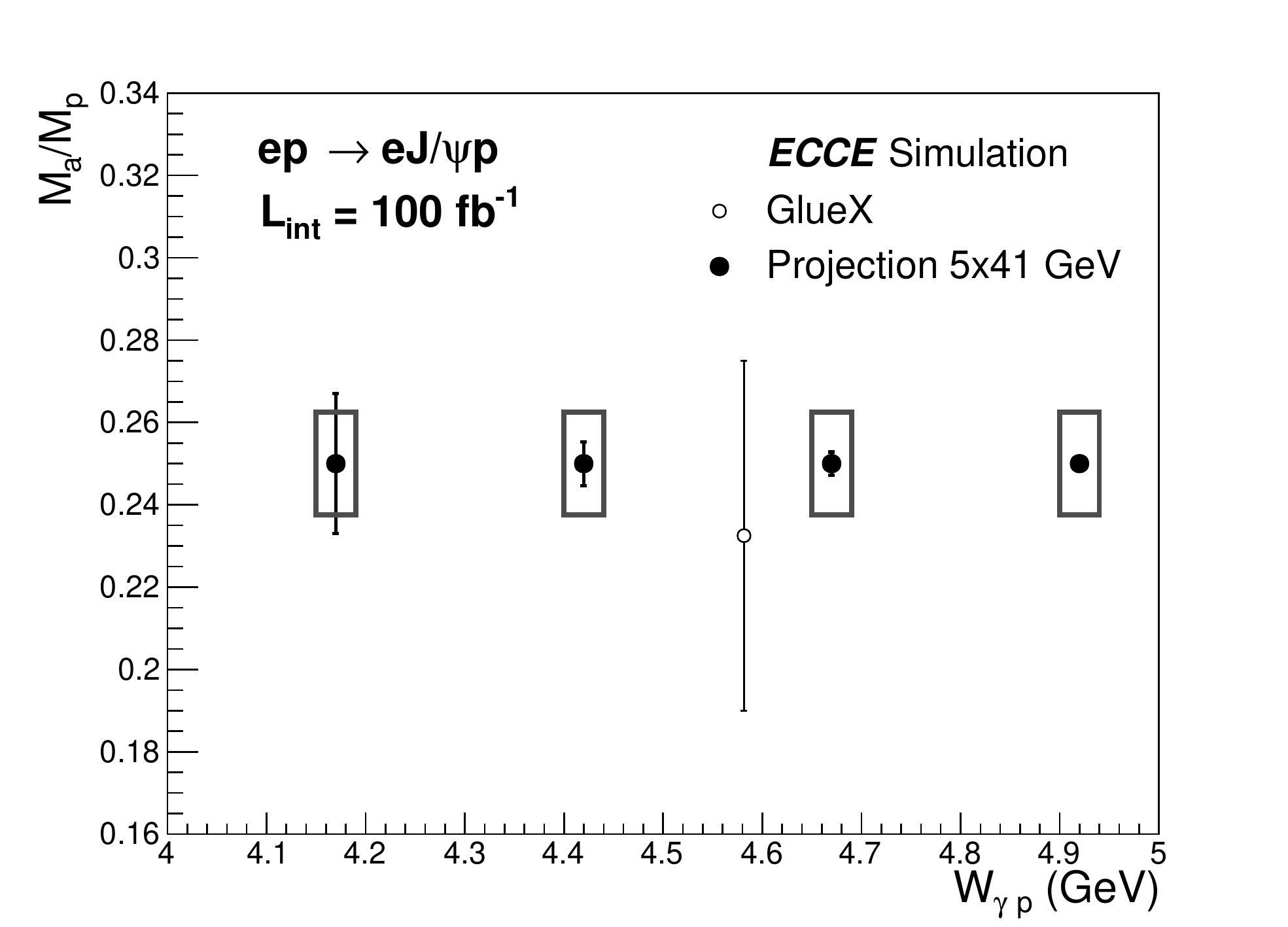}
    \caption{Trace anomaly contribution as a function of $\gamma$p center of mass energy}
    \label{fig:ano}
\end{figure}

\begin{figure}
    \centering
    \includegraphics[width=0.8\linewidth]{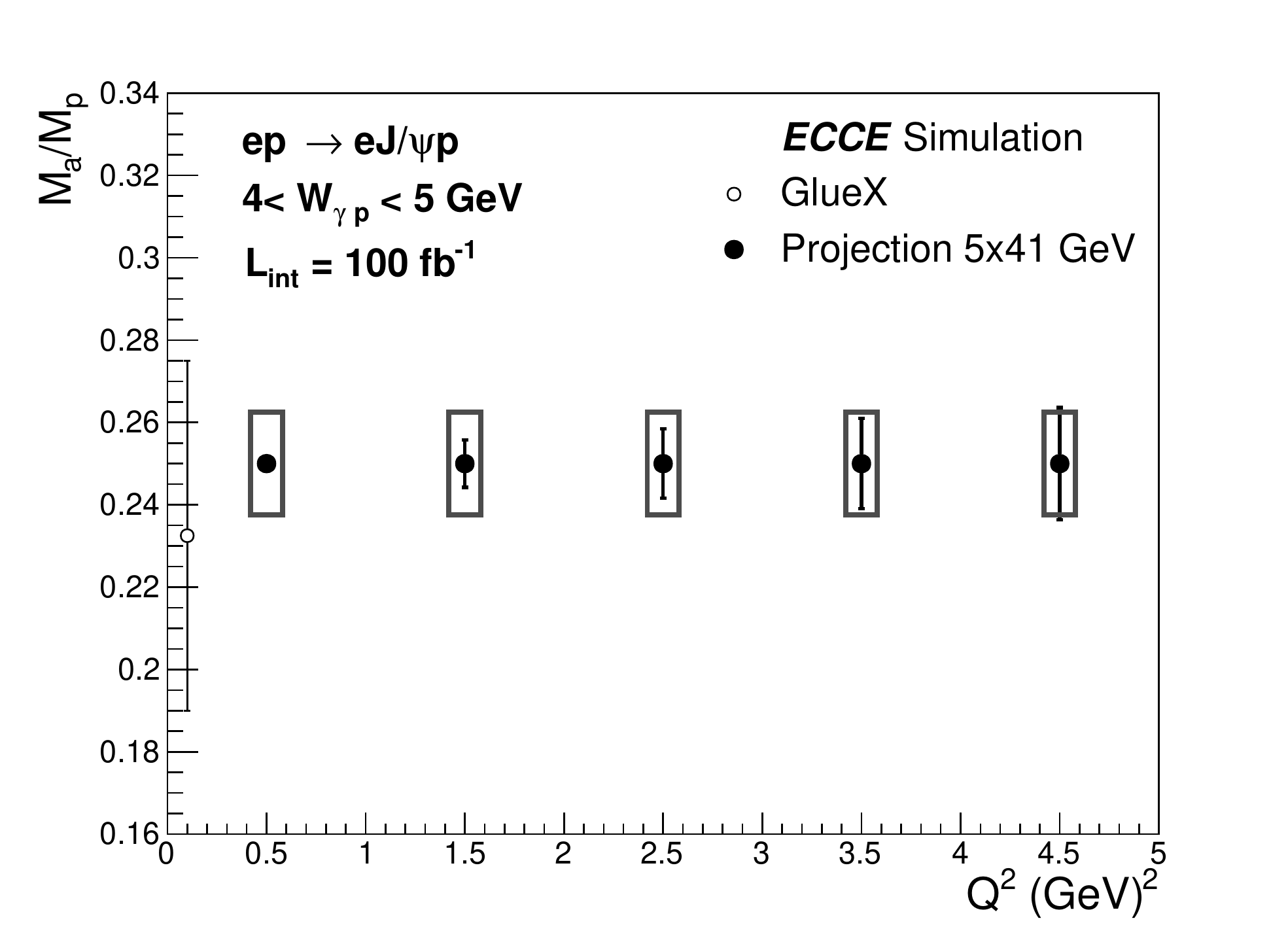}
    \caption{Trace anomaly contribution as a function of $Q^2$}
    \label{fig:ma}
\end{figure}

Recent theoretical efforts from VMD model and Holographic model~\cite{PhysRevD.101.086003}\cite{PhysRevD.98.074003}\cite{VMD2020} suggest that the  trace anomaly parameter can be extracted by the near-threshold exclusive heavy quarkonia process via their production at $(d\sigma/dt)|_{t=t_{min}}$. In the simulation, we made the projection of the trace anomaly parameter restriction capability at ECCE. The results are shown in Fig.~\ref{fig:ano} and Fig.~\ref{fig:ma}, which can provide precise information on the nucleon mass decomposition. The GlueX result~\cite{refId0} of the trace anomaly is also shown for comparison. The projection uncertainty consists of two parts, the statistical error using similar method as Sec.~\ref{g_dep} and systematical error defined in Sec.~\ref{t_dep}. The $d \sigma /dt|_{t=0}$ can be related to the trace anomaly parameter with Eq.(\ref{10},\ref{11},\ref{12})~\cite{refId0},  

\begin{equation}
\begin{aligned}
\label{10}
&\left.\frac{d \sigma_{\gamma N \rightarrow J / \psi N}}{d t}\right|_{t=0}= \left.\frac{3 \Gamma\left(J / \psi \rightarrow e^{+} e^{-}\right)}{\alpha m_{J / \psi}}\left(\frac{k_{J / \psi N}}{k_{\gamma N}}\right)^{2} \frac{d \sigma_{J / \psi N \rightarrow J / \psi N}}{d t}\right|_{t=0},
\end{aligned}
\end{equation}

\begin{equation}
\label{11}
\left.\frac{d \sigma_{J / \psi N \rightarrow J / \psi N}}{d t}\right|_{t=0}=\frac{1}{64 \pi} \frac{1}{m_{J / \psi}^{2}\left(\lambda^{2}-m_{N}^{2}\right)}\left|F_{J / \psi N}\right|^{2},
\end{equation}

where $k_{ab}^{2}=\left[s-\left(m_{a}+m_{b}\right)^{2}\right]\left[s-\left(m_{a}-m_{b}\right)^{2}\right] / 4 s$ denotes the squared momentum of center-of-mass of the corresponding two-body system, $\Gamma$ is the decay width of specific channel, $\alpha$ is the fine structure constant. $\lambda=\left(p_{N} p_{J / \psi} / m_{J / \psi}\right)$ is the energy of nucleon in the $J / \psi$ rest frame. At low energy, the forward amplitude $F_{J / \psi N}$ can be approximately written as a function of (1-b) in Eq.(\ref{12}), and the relative uncertainty of $d \sigma / dt|_{t=0}$ can be used to get the uncertainty of $M_a/M_p$ ($\propto (1-b)$) via the error propagation formula. 

\begin{equation}
\begin{aligned}
\label{12}
F_{J / \psi N} & \simeq r_{0}^{3} d_{2} \frac{2 \pi^{2}}{27}\left(2 M_{N}^{2}-\left\langle N\left|\sum_{i=u, d, s} m_{i} \bar{q}_{i} q_{i}\right| N\right\rangle\right) \\
& \simeq r_{0}^{3} d_{2} \frac{2 \pi^{2}}{27}\left(2 M_{N}^{2}-2 b M_{N}^{2}\right) \\
& \simeq r_{0}^{3} d_{2} \frac{2 \pi^{2}}{27} 2 M_{N}^{2}(1-b),
\end{aligned}
\end{equation}
where $r_0$ is the "Bohr" radius of $J/\psi$, and $d_2$ is the Wilson coefficient. These two parameters can be treated as constant in the relationship between $d \sigma / dt|_{t=0}$ and $(1-b)$ at low energy, thus could be neglected in the uncertainty determination.
\section{Summary}
\label{sum}
In this paper, we simulate exclusive J/$\psi$ production using Fun4All framework with the designed ECCE detector system at the future EIC. For J/$\psi$ detection, ECCE has good reconstruction efficiency and broad coverage, with large statistics for the designed EIC luminosity. We also demonstrate the capability of ECCE to probe the related physics opportunities for the exclusive J/$\psi$ photoproduction process. For gluon distribution in the proton and nucleus, the projection of the gluon nuclear shadowing effect shows an excellent capability of constraining the nuclear gluon PDF with the exclusive J/$\psi$ forward scattering measurements at ECCE. Benefited from the unprecedented coverage and excellent reconstruction capability, ECCE can provide a strong constraint to the near-threshold production mechanism of the exclusive J/$\psi$ photoproduction process. Furthermore, the projection results of the near-threshold exclusive heavy quarkonia production also show an excellent capability to extract the trace anomaly parameter to precisely determine the nucleon mass decomposition.  
\section*{Acknowledgement}

X. Li and W. Zha are supported by the National Natural Science Foundation of China (12005220, 12175223) and MOST(2018YFE0104900). The authors would like to thank the ECCE Consortium for performing a full simulation of their detector design, for providing up-to-date information on EIC run conditions, and for suggestions and comments on the manuscript. X. Li and W. Zha would like to thank Y. Zhou for useful suggestions and discussions related to this analysis.

\bibliographystyle{elsarticle-num} 
\bibliography{main}

\end{document}